\documentclass[letterpaper]{sig-alternate-2013}

\permission{\copyright 2017 International World Wide Web Conference Committee (IW3C2),
published under Creative Commons CC BY 4.0 License.}
\conferenceinfo{WWW 2017,}{April 3--7, 2017, Perth, Australia.}
\copyrightetc{ACM \the\acmcopyr}
\crdata{978-1-4503-4913-0/17/04 \\
http://dx.doi.org/10.1145/3038912.3052677 \\
\includegraphics{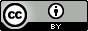}
}

\usepackage{graphicx}
\usepackage{mathptmx}
\usepackage{amssymb}
\usepackage{amsmath}
\usepackage{subfigure}
\usepackage{hyphenat}
\usepackage{color}
\usepackage{textcomp}
\usepackage{enumitem}
\usepackage{tabularx}
\usepackage{mathtools}
\usepackage{ucs}
\usepackage[utf8x]{inputenc}
\usepackage[hyphens]{url}

\usepackage{balance}

\newcommand{\hide}[1]{}
\newcommand{\xhdr}[1]{\vspace{1.7mm}\noindent{{\bf #1.}}}
\newcommand{\xhdrNoPeriod}[1]{\vspace{1.7mm}\noindent{{\bf #1}}}

\newcommand{\ie}{\textit{i.e.}}
\newcommand{\eg}{\textit{e.g.}}

\begin{document}

\title{
An Army of Me: Sockpuppets in Online Discussion Communities
}

\numberofauthors{4}
\author{
\alignauthor Srijan Kumar\\
\affaddr{University of Maryland}\\
\email{srijan@cs.umd.edu}
\alignauthor Justin Cheng\\
\affaddr{Stanford University}\\
\email{jcccf@cs.stanford.edu}
\and
\alignauthor Jure Leskovec\\
\affaddr{Stanford University}\\
\email{jure@cs.stanford.edu}
\alignauthor V.S. Subrahmanian\\
\affaddr{University of Maryland}\\
\email{vs@cs.umd.edu}
}
\maketitle

\vspace{-3mm}

\begin{abstract}

In online discussion communities, users can interact and share information and opinions on a wide variety of topics.
However, some users may create multiple identities, or sockpuppets, and engage in undesired behavior by deceiving others or manipulating discussions.
In this work, we study sockpuppetry across nine discussion communities, and show that sockpuppets differ from ordinary users in terms of their posting behavior, linguistic traits, as well as social network structure.
Sockpuppets tend to start fewer discussions, write shorter posts, use more personal pronouns such as ``I'', and have more clustered ego-networks.
Further, pairs of sockpuppets controlled by the same individual are more likely to interact on the same discussion at the same time than pairs of ordinary users.
Our analysis suggests a taxonomy of deceptive behavior in discussion communities.
Pairs of sockpuppets can vary in their deceptiveness, i.e., whether they pretend to be different users, or their supportiveness, i.e., if they support arguments of other sockpuppets controlled by the same user. 
We apply these findings to a series of prediction tasks, notably, to identify whether a pair of accounts belongs to the same underlying user or not.
Altogether, this work presents a data-driven view of deception in online discussion communities and paves the way towards the automatic detection of sockpuppets.

\end{abstract}

\vspace{-3mm}

\vspace{-2mm}

\section{Introduction}

\begin{figure}[t]
    \centering
    \hspace{-3mm}
    \includegraphics[width=0.8\columnwidth, clip=true, trim=15mm 15mm 12mm 15mm]{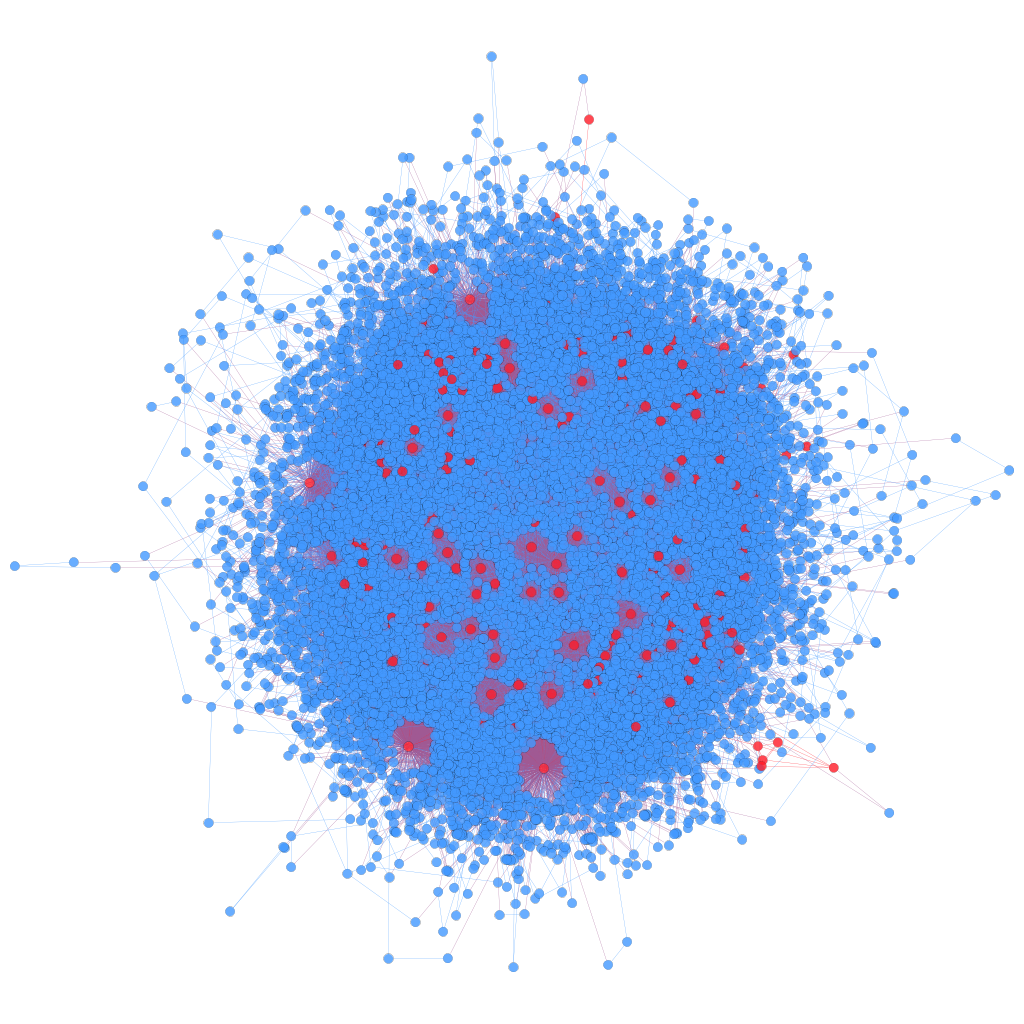}
    \caption{AVClub.com social network. Nodes represent users and edges connect users that reply to each other. Sockpuppets (red nodes) tend to interact with other sockpuppets, and are more central in the network than ordinary users (blue nodes).}
    \label{fig:avclub}
    \vspace{-5mm}
\end{figure}

Discussions are a core mechanism through which people interact with one another and exchange information, ideas, and opinions.
They take place on social networks such as Facebook, news aggregation sites such as Reddit, as well as news websites such as CNN.com.
Nonetheless, the anonymity afforded by some discussion platforms has led to some users deceiving others using multiple accounts, or sockpuppets~\cite{donath1999identity}.
Sockpuppetry is often malicious and deceptive, and has been used to manipulate public opinion \cite{afroz2012detecting, stone2007hand} and vandalize content (e.g., on Wikipedia \cite{solorio2013case}).

Prior work on sockpuppetry and deceptive behavior has tended to focus on individual motivations \cite{caspi2006online,hancock2007digital}, or on identifying sockpuppets through their linguistic traits \cite{bu2013sock} (e.g., on Wikipedia \cite{solorio2013case,tsikerdekis2014multiple}).
Further, given the difficulty of obtaining ground-truth data about sockpuppets, work has also tended to make assumptions about how sockpuppets behave, for example, assuming that they have similar usernames \cite{liu2016sockpuppet}, they are only used to support one another \cite{zheng2011sockpuppet}, or that they write similar to each other \cite{bu2013sock}.
Further, research has generally not considered how the interactions between sockpuppets controlled by the same individual could be used to accurately and automatically identify sockpuppets.
As such, improved methods for identifying sockpuppets, as well as deeper analyses of how sockpuppets interact with one another may allow us to better understand, characterize, and automatically detect sockpuppetry.

\xhdr{The present work: Sockpuppetry in online discussion communities}
In this paper, we focus on identifying, characterizing, and predicting sockpuppetry in nine different online discussion communities.
We broadly define a \emph{sockpuppet} as a user account that is controlled by an individual (or {\em puppetmaster}) who controls at least one other user account.
By considering less easily manipulated behavioral traces such as IP addresses and user session data, we automatically identified 3,656 sockpuppets comprising 1,623 \emph{sockpuppet groups}, where a group of sockpuppets is controlled by a single puppetmaster.

Studying these identified sockpuppets, we discover that sockpuppets differ from ordinary users in terms of how they write and interact with other sockpuppets.
Sockpuppets have unique linguistic traits, for example, using more singular first-person pronouns (e.g., ``I''), corroborating with prior work on deception~\cite{afroz2012detecting}.
They also use fewer negation words, perhaps in an attempt to appear more impartial, as well as fewer standard English parts-of-speech such as verbs and conjunctions.
Suggesting that sockpuppets write worse than ordinary users on average, we find that posts are more likely to be downvoted, reported by the community, and deleted by moderators.
Sockpuppets also start fewer discussions.

Examining pairs of sockpuppets controlled by the same puppetmaster, we find that sockpuppets are more likely to post at the same time and post in the same discussion than random pairs of ordinary users.
As illustrated in Figure~\ref{fig:avclub}, by studying the network of user replies, we find that sockpuppets have a higher pagerank and higher local clustering coefficient than ordinary users, suggesting that they are more important in the network and tend to generate more communication between their neighbors.
Further, we find that pairs of sockpuppets write more similarly to each other than to ordinary users, suggesting that puppetmasters tend not to have both ``good'' and ``bad'' accounts.

While prior work characterizes sockpuppetry as malicious~\cite{solorio2013case, gani2012towards, tsikerdekis2014multiple}, not all the sockpuppets we identified were malicious.
In some sockpuppet groups, sockpuppets have display names significantly different from each other, but in other groups, they have more similar display names.
Our findings suggest a dichotomy in how deceptive sockpuppets are -- some are {\em pretenders}, that masquerade as separate users, while others are {\em non-pretenders}, that is sockpuppets that are overtly visible to other members of the community.
Pretenders tend to post in the same discussions and are more likely to have their posts downvoted, reported, or deleted compared to non-pretenders.
In contrast, non-pretenders tend to post in separate discussions, and write posts that are longer and more readable.

Our analyses also suggest that sockpuppets may differ in their supportiveness of each other.
Pairs of sockpuppets controlled by the same puppetmaster differ in whether they agree with each other in a discussion.
While sockpuppets in a pair mostly remain neutral towards each other (or are {\em non-supporters}), 30\% of the time, one sockpuppet in a pair is used to support the other (or is a {\em supporter}), while 10\% of the time, one sockpuppet is used to attack the other (or is a {\em dissenter}).
Studying both deceptiveness and supportiveness, we find that supporters also tend to be pretenders, but dissenters are not more likely to be pretenders, suggesting that deceptiveness is only important when sockpuppets are trying to create an illusion of public consensus.

Finally, we show how our previous observations can be used to develop models for automatically identifying sockpuppetry.
We demonstrate robust performance in differentiating pairs of sockpuppets from pairs of ordinary users (ROC AUC=0.90), as well as in the more difficult task of predicting whether an individual user account is a sockpuppet (ROC AUC=0.68).
We discover that the strongest predictors of sockpuppetry relate to interactions between the two sockpuppet accounts, as well as the interactions between the sockpuppet and the community.

Altogether, our results begin to shed light on how sockpuppetry occurs in practice, and paves the way towards the development and maintenance of healthier online communities.

\newpage

\section{Data and Definitions}
We start by laying out the terminology that we use in the remainder of the paper.
We then describe the data we used in our analysis and a robust method for automatically identifying sockpuppets.

\xhdr{Sockpuppetry}
While in prior work sockpuppets have typically been used to refer to a false online identity that is used for the purposes of deceiving others~\cite{liu2016sockpuppet, zheng2011sockpuppet}, we adopt a broader definition.
We define a \emph{sockpuppet} as a user account controlled by an individual who has at least one other account.
In other words, if an individual controls multiple user accounts, each account is a sockpuppet.
The individual who controls these sockpuppets is referred to as the \emph{puppetmaster}.
We use the term \emph{sockpuppet group/pair} to refer to all the sockpuppets controlled by a single puppetmaster. 
In each sockpuppet group, one sockpuppet typically has made significantly more comments than the others -- we refer to this sockpuppet as the \emph{primary} sockpuppet, and the other sockpuppets as \emph{secondary}.
Finally, we use \emph{ordinary user} to refer to any user account that is not a sockpuppet.

We study sockpuppets in the context of online discussion communities.
In these communities, people can create accounts to comment on articles.
In addition to writing or replying to \emph{posts}, users can also vote on posts or report them for abuse.
Moderators can, in turn, delete posts that do not conform to community standards.
If a post is not a reply to another post, we call that post a \emph{root post}.
We define a \emph{discussion} as all the posts that follow a given news article, and a \emph{sub-discussion} as a root post and any replies to that post.

\begin{table}[t!]
\small
\begin{tabular}{llccc}
\hline
Community & Genre & \# Users & \# Sockpuppets & \# Sock-groups\\
\hline
CNN & News & 846,436 & 1,191 & 523\\
Breitbart & News & 196,846 & 761 & 352\\
allkpop & Music & 159,671 & 445 & 193\\
MLB & Sports & 115,845 & 339 & 139\\
IGN & Games & 266,976 & 314 & 142\\ 
Fox News & News & 145,009 & 214 & 94\\
A.V. Club & Entertainment & 37,332 & 199 & 90\\
The Hill & Politics & 158,378  & 134 & 62\\
NPR & News & 65,662  & 59 & 28\\
\hline
\end{tabular}
\vspace{-3mm}
\caption{Statistics of the nine online discussion communities.}
\vspace{-3mm}
\label{tab:stats}
\end{table}

\xhdr{Data}
The data consists of nine different online discussion communities that encompass a variety of topical interests -- from news and politics to sports and entertainment (Table~\ref{tab:stats}).
Disqus, a commenting platform that hosted these discussions, provided us with a complete trace of user activity across nine communities that consisted of 2,897,847 users, 2,129,355 discussions, and 62,744,175 posts.
Each user has a display name which appears next to their posts and an email address which is private. (To respect user privacy, all email addresses were stripped of the domain names and analyzed in aggregate.)
Each post is also associated with an anonymized IP address of the posting user.

\xhdr{Identifying sockpuppets}
No explicit labels of sockpuppets exist in any of the discussion communities, so to identify sockpuppets, we use multiple signals that together suggest that accounts are likely to share the same owner -- the IP address of a comment, as well as the times at which comments are made, and the discussions they post in.
Our approach draws on the approach adopted by Wikipedia administrators who identify sockpuppets by finding accounts that make similar edits on the same Wikipedia article, in near-similar time and from same IP address~\cite{wikisocks}.
As we are primarily interested in identifying sockpuppets with high precision, the criteria we adopt is relatively conservative -- relaxing these criteria may improve recall, but at the cost of more false positives.

To limit spurious detection of sockpuppets, we filter the data and remove any IP addresses used by many user accounts, as these accounts may simply be accessed from behind a country-wide proxy or intranet. 
We also do not consider user accounts that post from many different IP addresses, since they have high chance of sharing IP address with other accounts.
Specifically, for each discussion community, we remove the top 5\% most used IP addresses and the top 5\% accounts that have the most IP addresses.

Then, we identify sockpuppets as user accounts that post from the same IP address in the same discussion within $T$ minutes for at least $K_{min}$ different discussions. 
Here, we set $T =$ 15 minutes (larger values result in empirically similar findings).
To pick an appropriate value for $K_{min}$, we use two metrics that prior work has found indicative of sockpuppets: the time difference between posts made by two accounts, and the difference in the length of posts~\cite{bu2013sock, tsikerdekis2014multiple, johansson2013detecting, qian2013identifying}.

\begin{figure} [t!]
    \centering
    \hspace{-3mm}
    \subfigure{
        \includegraphics[width=0.6\columnwidth, trim=3mm 2.4cm 3mm 3mm]{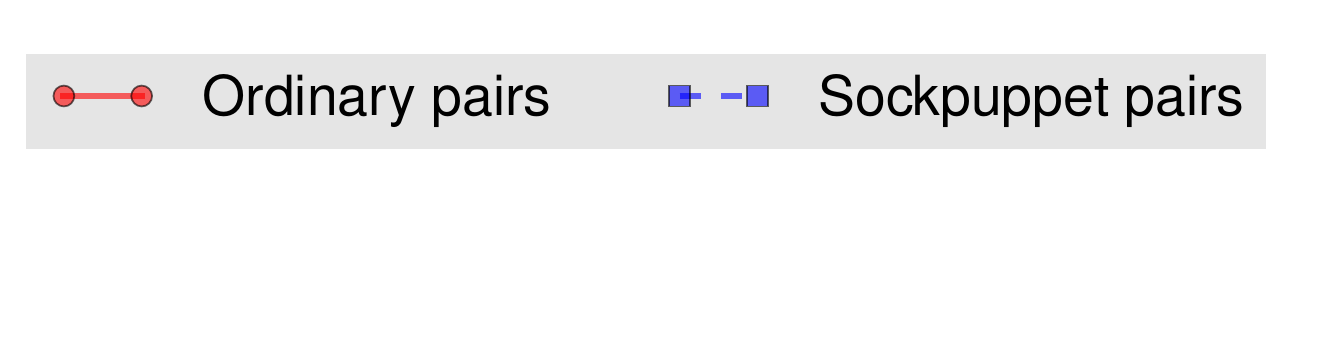}
    }
    
    \addtocounter{subfigure}{-1}
    \hspace{-3mm}
    \subfigure[\hspace{-9mm}]{
        \includegraphics[width=0.49\columnwidth]{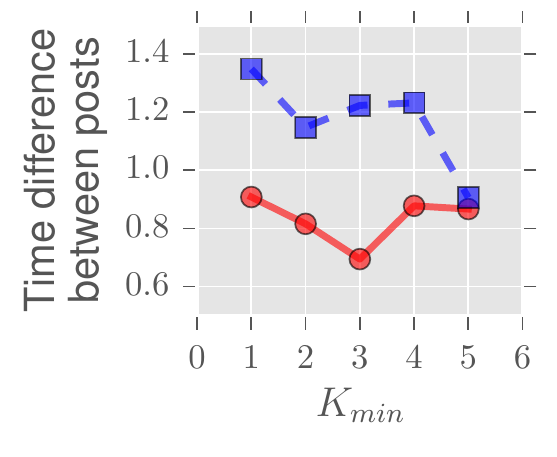}
        \label{fig:cluster_sizes}
    }
    \hspace{-3mm}
    \subfigure[\hspace{-9mm}]{
        \includegraphics[width=0.49\columnwidth]{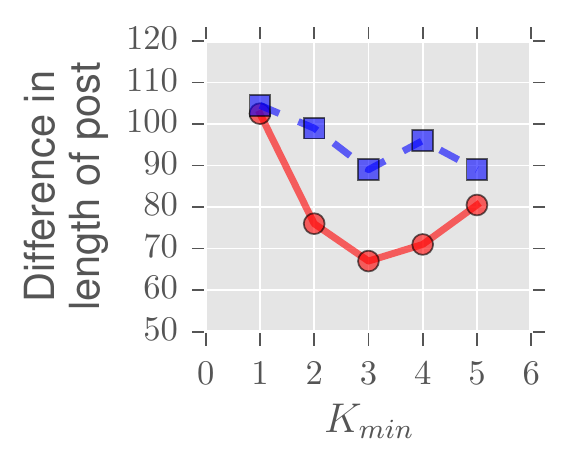}
        \label{fig:cluster_sizes}
    }

    \vspace{-3mm}
    \caption{Varying $K_{min}$, the minimum number of overlapping sessions between users for them to be identified as sockpuppets. For sockpuppet pairs (blue) the time between posts, and the difference in post lengths reach a minimum value at $K_{min}=3$.}
    \label{fig:kvar}
    \vspace{-5mm}
\end{figure} 

Figure \ref{fig:kvar} plots these quantities for identified pairs of sockpuppets as well as random pairs of users.
We observe that regardless of the value of $K_{min}$, the identified sockpuppets post more closely in time and write posts more similar in length, compared to a random pair of users.
Further, we find that among sockpuppet pairs, these quantities achieve their minimum at $K_{min} = 3$, which means that sockpuppets are most reliably identified at that value of $K_{min}$.

To summarize, we define sockpuppets as user accounts that post from the same IP address in the same discussion in close temporal proximity at least 3 times.
We then define sockpuppet groups as maximal sets of accounts such that each account satisfies the above definition with at least one other account in the group.
Overall, we identify a total of 1,623 sockpuppet groups, consisting of 3,656 sockpuppets from nine different online discussion communities (Table~\ref{tab:stats}).
As most sockpuppet groups contain two sockpuppets (Figure~\ref{fig:cluster_sizes}), we focus our analyses on pairs of sockpuppets.

We give an example of an identified sockpuppet group below, which consists of three users: $S_1$ and $S_2$ comprise a pair of sockpuppets, while $O$ is an ordinary user.
After an unusually positive interaction between the two sockpuppets, $O$ identifies them as being controlled by the same puppetmaster:

\vspace{-2mm}
{
\begin{quote}
$S_1$: Possibly the best blog I've ever read major props to you.

$\hookrightarrow$ $S_2$: Thanks.  I knew Marvel fans would try to flame me, but they have nothing other than ``oh that's your opinion'' instead of coming up with their own argument.

$\hookrightarrow$ $O$: Quit talking to yourself, *******. Get back on your meds if you're going to do that.
\end{quote} 
}

\vspace{-2mm}

\section{Characterizing Sockpuppetry}
Having identified sockpuppets, we now turn to characterizing their behavior. We study when sockpuppets are created and how their language and social networks differ from ordinary users across all nine discussion communities.

\begin{figure}[t]
    \centering
    \hspace{-3mm}
    \subfigure[\hspace{-9mm}]{
        \includegraphics[width=0.5\columnwidth]{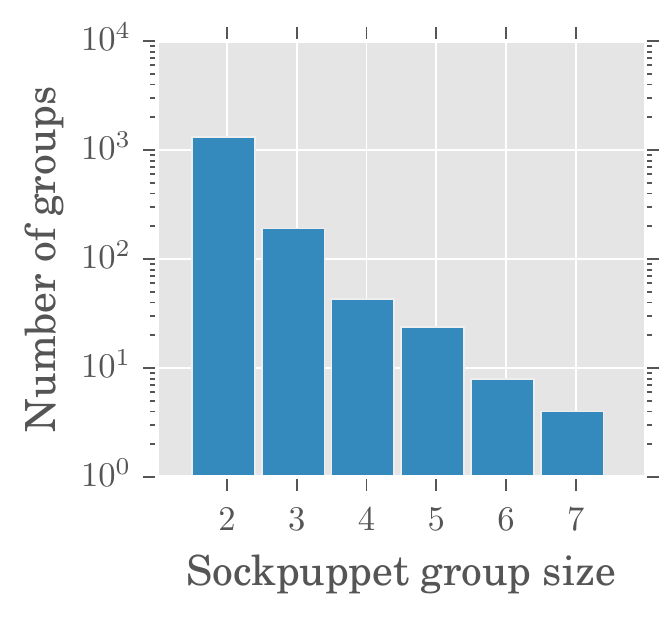}
        \label{fig:cluster_sizes}
    }
    \hspace{-3mm}
    \subfigure[\hspace{-9mm}]{
        \includegraphics[width=0.5\columnwidth]{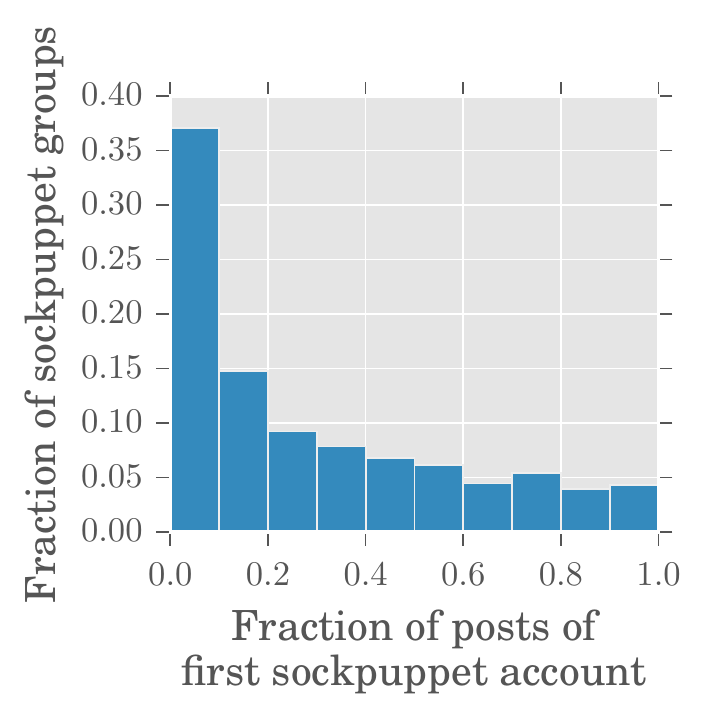}
        \label{fig:second_account_creation}
    }
    \vspace{-3mm}
    \caption{(a) Number of sockpuppet groups, \ie\ sockpuppets belonging to the same puppetmaster. (b) The second sockpuppet in a group tends to be created shortly after the first.}
\end{figure} 

\xhdr{Sockpuppets are created early}
To understand when sockpuppets are created, we examine the activity of the first sockpuppet account in each sockpuppet pair. 
Figure~\ref{fig:second_account_creation} shows the fraction of total number of posts made by the first sockpuppet before the second sockpuppet is created.
The second sockpuppet tends to be created during the first 10\% of the posts, with a median of 18\% posts written by first sockpuppet before the second sockpuppet begins posting.
In other words, sockpuppets tend to be created early in a user's lifetime, which may indicate that sockpuppet creation is premeditated and not a result of user's interactions in the community.

\xhdr{Matching sockpuppets with ordinary users}
On average, sockpuppets write more posts than ordinary users (699 vs. 19) and participate in more discussions (141 vs. 7).
To control for this disparity, in all our subsequent analyses we use propensity score matching~\cite{rosenbaum1983central} to match sockpuppets with ordinary users that have similar numbers of posts and make posts to the same set of discussions.

\begin{figure}[t]
    \centering
    \includegraphics[width=0.35\textwidth]{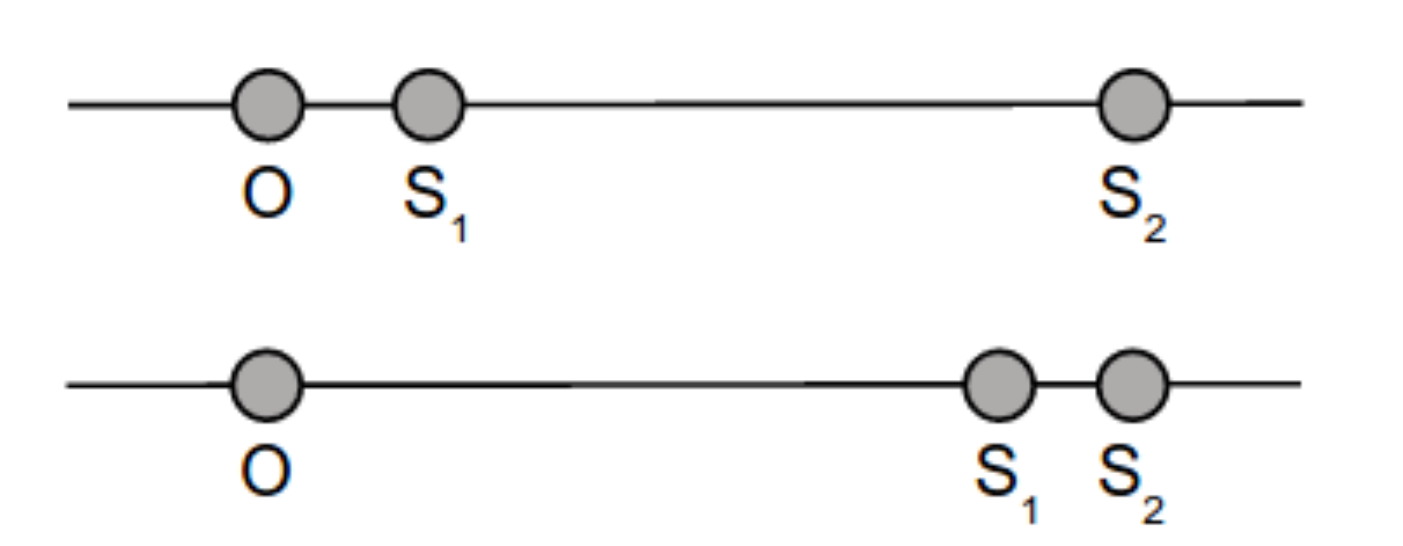}
    \vspace{-2mm}
    \caption{Two hypotheses how similarity of sockpuppet pairs and ordinary users relates to each other. Top: Under the double life hypothesis, sockpuppet $S_1$ is similar to an ordinary user $O$, while $S_2$ deviates. Bottom: Alternative hypothesis is that both sockpuppet accounts are highly different from ordinary users.}
    \label{fig:regime}
    \vspace{-3mm}
\end{figure}

\begin{figure*}[t]
    \centering
    \hspace{-3mm}
    \subfigure[\hspace{-9mm}]{
        \includegraphics[width=0.2\textwidth]{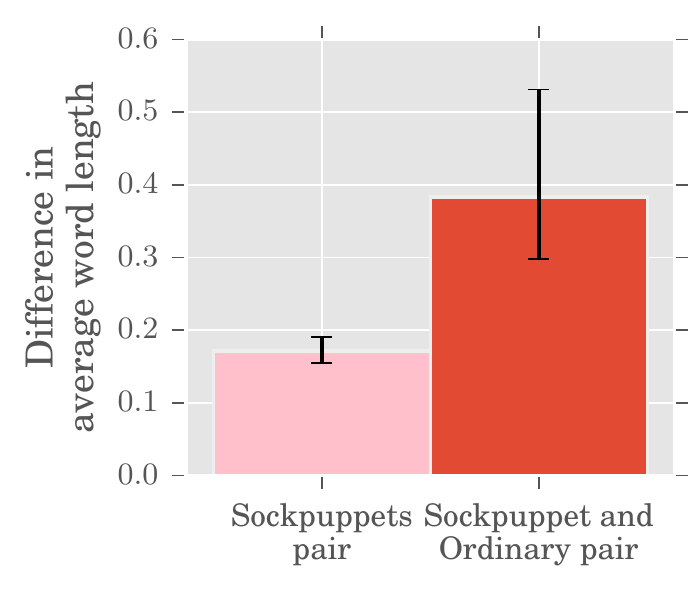}
        \label{fig:ssvsr-avg_word_length}
    }
    \hspace{-3mm}
    \subfigure[\hspace{-9mm}]{
        \includegraphics[width=0.2\textwidth]{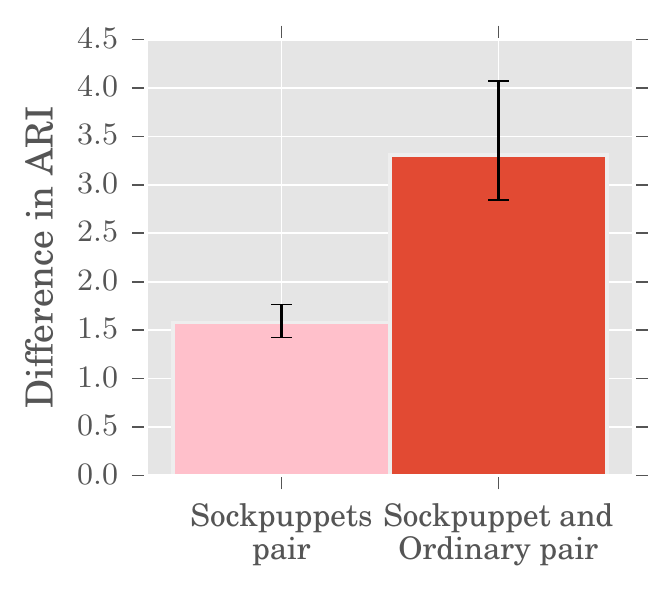}
        \label{fig:ssvsr-ARI}
    }
    \hspace{-3mm}
    \subfigure[\hspace{-9mm}]{
        \includegraphics[width=0.2\textwidth]{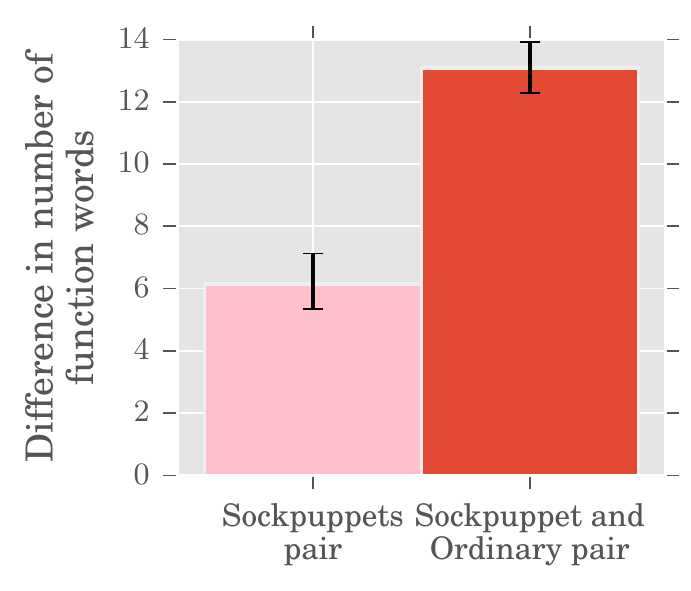}
        \label{fig:ssvsr-funct}
    }
    \hspace{-3mm}
    \subfigure[\hspace{-9mm}]{
        \includegraphics[width=0.2\textwidth]{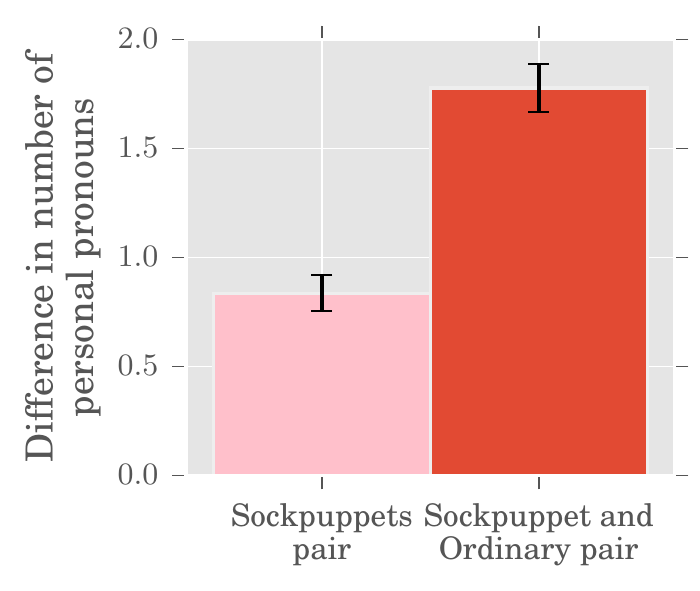}
        \label{fig:ssvsr-ppron}
    }
    \hspace{-3mm}
    \subfigure[\hspace{-9mm}]{
        \includegraphics[width=0.2\textwidth]{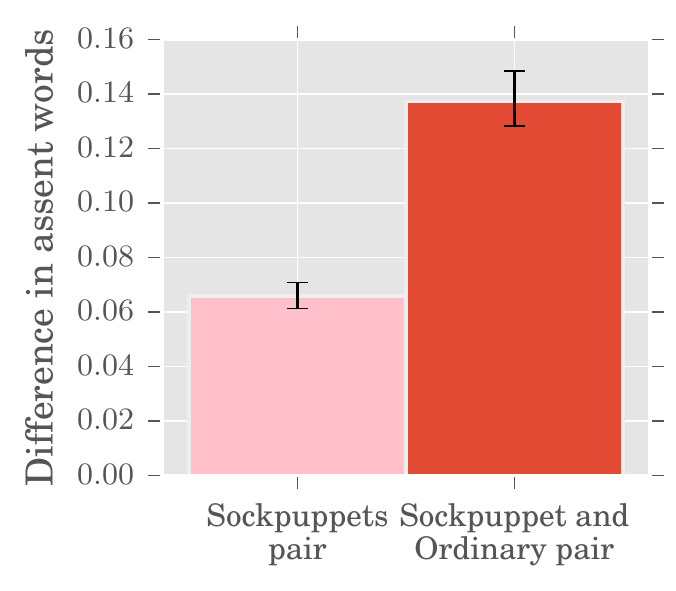}
        \label{fig:ssvsr-assent}
    }
    \vspace{-3mm}
    \caption{Difference in properties of sockpuppet pairs and that of sockpuppet-ordinary pairs. Sockpuppet pairs are more similar to each other in several linguistic attributes.}
    \label{fig:doublelife}
    \vspace{-4mm}
\end{figure*}

\subsection{Do puppetmasters lead double lives?}
First, we explore an important question about how the behavior of sockpuppets $S_1$ and $S_2$ controlled by the same puppetmaster relates to that of an ordinary user $O$.
Two possible hypotheses are illustrated in Figure~\ref{fig:regime}.
First is the \emph{double life} hypothesis, where the puppetmaster maintains a distinct personality for each sockpuppet -- one sockpuppet, $S_1$, behaves like an ordinary user, while the other, $S_2$ behaves more maliciously.
Under this hypothesis we would expect that the linguistic similarity of posts written by $O$ and $S_1$ would be high, and that between both $S_1$ and $S_2$, as well as $O$ and $S_2$ to be significantly lower.
In the alternative hypothesis (Figure~\ref{fig:regime} (bottom)), both sockpuppets act maliciously.
In this case, we might expect that the linguistic similarity between $S_1$ and $S_2$ would be low, but that between $S_1$ and $O$, and $S_2$ and $O$ to be much lower.

To find out which is the case, we compare the language of pairs of sockpuppets, and that of each sockpuppet with an ordinary user.
To control for user activity, we again match sockpuppets with ordinary users that have similar posting activity, and that participate in similar discussions. Specifically, for each user, we created a feature vector consisting of several linguistic features computed from that user's posts, including LIWC categories and sentiment, the average number of words in a post, the average fraction of special characters. We then compute the cosine similarity of the feature vectors.

We find that on average, the two sockpuppets are more similar to each other than either is to an ordinary user ($p$ < 0.001). Figure~\ref{fig:doublelife} highlights that these observations hold for individual features as well -- the difference between two sockpuppets' readability score (ARI), average word length, number of function words, personal pronouns and assent words are smaller than that of either sockpuppet and an ordinary user.
This suggests that the double life hypothesis (Figure~\ref{fig:regime}(top)) is less likely to be true than the alternate hypothesis (Figure~\ref{fig:regime}(bottom)).

In other words, this experiment suggests that puppetmasters do not lead double lives, and that it is generally not the case that individual sockpuppets controlled by the same puppetmaster behave differently.
Rather, sockpuppets as a whole tend to write differently from ordinary users, and sockpuppets controlled by the same puppetmaster all tend to write similarly to each other.

\vspace{-1.5mm}
\subsection{Linguistic Traits of Sockpuppets}
Having established that different sockpuppets controlled by the same puppetmaster behave consistently, we now turn our attention to  quantify their linguistic traits more precisely.
Here, we focus on comparing various measures of similarity $sim(S_i, O)$ of a sockpuppet $S_i$ ($i=\{1,2\}$) and a matched ordinary user $O$.
Specifically, we use LIWC word categories~\cite{pennebaker2001linguistic} to measure the fraction of each type of words written in all posts, and VADER~\cite{vader} to measure sentiment of posts.
We report the average values for sockpuppets and the corresponding $p$-values by performing paired $t$-tests for each sockpuppet and its matching ordinary user.

\xhdrNoPeriod{Do sockpuppets write differently from ordinary users?}
Linguistic traits have been used to identify trolls in discussions~\cite{cheng2015antisocial}, vandalism on Wikipedia~\cite{potthast2008automatic}, and fake reviewers on e-commerce platforms~\cite{mukherjee2013yelp}. For example, deceptive authors tend to increase their usage of function words, particles and personal pronouns~\cite{afroz2012detecting, brennan2012adversarial}.
They use more first- and second-person singular personal pronouns (e.g., `I', `you'), while reducing their use of third-person singular personal pronouns (e.g., `his', `her').
They also tend to oversimplify their writing style by writing shorter sentences and writing words with fewer syllables.

We make similar observations with respect to sockpuppets.
First, we observe that they tend to write posts that are more self-centered -- and use ``I'' more often than ordinary users (0.076 for sockpuppets vs 0.074 for ordinary users, $p$<0.001).
Sockpuppets also use ``you'' more often (0.017 vs 0.015, $p$<0.01) but third-person singular personal pronouns and plural personal pronouns (i.e., `we', `he', `she', and `they') less (\eg\ 0.016 vs 0.018 for `he/she' words, $p$ < 0.001), indicating that they tend to address other users in the community more directly.
Similarly, we observe that sockpuppets also write shorter sentences than ordinary users (a mean of 12.4 vs. 12.9 words per sentence, $p$ < 0.001).
However, in contrast to prior work on deceptive writing, sockpuppets use a similar number of syllables per word (1.29 vs 1.28, $p$ = 0.35).


Turning to differences in LIWC categories, we observe that sockpuppets also appear to write worse than ordinary users.
They are more likely to swear (0.003 vs 0.002, $p$ < 0.05) and use more punctuation (0.057 vs 0.055, $p$ < 0.05), while using fewer alphabetic characters (0.769 vs 0.771, $p$ < 0.05).
Sockpuppets also use fewer standard English parts-of-speech, such as articles, verbs, adverbs and conjuctions (\eg\ for articles, 0.062 vs 0.064, $p$ < 0.001).
However, while trolls wrote posts that were less readable \cite{cheng2015antisocial}, sockpuppets write posts with similar readability (automated readability index, or ARI = 11.24 vs 11.41 of ordinary users, $p$=0.09).
Sockpuppets also tend to agree more in their posts (0.002 vs -0.012, $p$ < 0.05), possibly to minimize conflict with others and support their other sockpuppet account.
They also express less negative sentiment (0.022 vs 0.023, $p$ < 0.001), though their overal sentiment, subtracting negative from positive sentiment, is similar to that of ordinary users (0.030 vs 0.028, $p$ = 0.43).

\begin{figure}
    \centering
    \hspace{-3mm}
    \subfigure[\hspace{-9mm}]{
        \includegraphics[width=0.5\columnwidth]{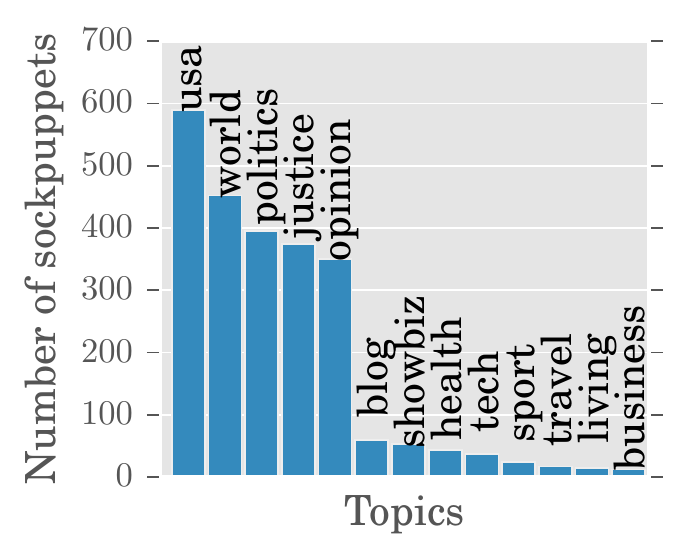}
        \label{fig:most-active-topics}
    }
    \hspace{-3mm}
    \subfigure[\hspace{-9mm}]{
        \includegraphics[width=0.5\columnwidth]{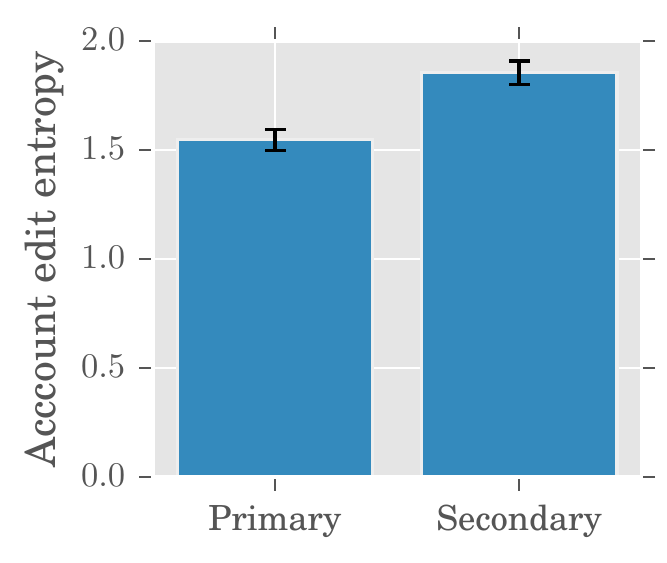}
        \label{fig:entropy}
    }
    \vspace{-3mm}
    \caption{
    (a) Histogram for the most active topic for each sockpuppet account.
    (b) In a sockpuppet group, the secondary sockpuppets tend to be used alongside the primary sockpuppet.
    }
    \vspace{-6mm}
\end{figure} 

\vspace{-2mm}
\subsection{Activity and Interactions}
Next, we study how sockpuppets interact with the community at large, and how it responds to these sockpuppets.

\xhdr{Sockpuppets start fewer discussions, and post more in existing discussions}
First, we note that sockpuppets start fewer discussions, but rather post more within existing discussions (65\% of sockpuppets' posts are replies compared to 51\% for ordinary users, $p$ < 0.001).
This shows that sockpuppets are mainly used to reply to other users. 

\xhdr{Sockpuppets tend to participate in discussions with more controversial topics}
Do sockpuppets create accounts to participate in certain topics? 
To answer this, we look at the topics of the discussions in CNN on which sockpuppets post. As shown in Figure~\ref{fig:most-active-topics} topics that tend to attract more controversy such as \emph{usa, world, politics, justice} and \emph{opinion}, also attract the majority of sockpuppets, while other topics such as health and showbiz have comparatively fewer sockpuppets. This indicates that one of the main motivations for using sockpuppets is to use them to build support for a particular position, corroborating prior work~\cite{bu2013sock, zheng2011sockpuppet}.

\xhdr{Sockpuppets are treated harshly by the community}
A community can provide feedback to a sockpuppet in three ways: other users can vote on or report their posts, and moderators can delete the posts.
Comparing the posts made by sockpuppets with those made by ordinary users, we find that sockpuppets' posts receive a greater fraction of downvotes (0.51 vs 0.43, $p$ < 0.001), are reported more often (0.05 vs 0.026, $p$ < 0.001) and are also deleted more often (0.11 vs 0.08, $p$ < 0.001).
Moreover, sockpuppets are also blocked by the moderators more often (0.09 vs 0.07, $p$ < 0.001).
Overall, this suggests that sockpuppets are making undesirable comments.

\xhdr{Sockpuppets in a pair interact with each other more}
Pairs of sockpuppets also tend to post together in more sub-discussions compared to random pairs of ordinary users (6.57 vs 0.33, $p$ < 0.001).
Moreover, looking at when posts are made, pair of sockpuppets also post more frequently on the same discussion within 15 minutes of each other (7.8 vs 4.28, $p$ < 0.001). In other words, pairs of sockpuppets are significantly more likely to interact with one another, and post at the same time, than two ordinary users would.

\xhdr{Sockpuppets in a pair upvote each other more}
Looking at votes, pairs of sockpuppets vote significantly more often on each other's posts than random pairs of ordinary users (9.35 vs 0.40 votes, $p$ < 0.001).
Among the two sockpuppets in a pair, the secondary sockpuppet votes more on primary sockpuppet's posts than vice-versa (14.2 vs 4.5 votes, $p$ < 0.01).
Moreover, pair of sockpuppets largely give positive votes to each other as compared to ordinary users (0.987 vs 0.952; $p$ < 0.05). 
Altogether, sockpuppets in a pair use their votes to significantly inflate one another's `popularity'.

\xhdr{Secondary sockpuppets are used in conjunction with primary sockpuppets}
Puppetmasters, while controlling multiple sockpuppets, may either use multiple sockpuppets at the same time, or different sockpuppets at different times.
To quantify how a puppetmaster may switch between using different sockpuppets, we compute the fraction of consecutive posts made by a particular sockpuppet, and then compute the entropy of this distribution. This way we quantify how much intertwined is the usage of both sockpuppet accounts.
For instance, consider two sockpuppets controlled by the same puppetmaster.
The puppetmaster first uses $S_1$ to write 5 posts, then uses $S_2$ to write 1 post, switches back to $S_1$ to write 4 more posts, and then finally switches back to $S_2$ to write 1 more post.
The entropy of this post sequence for $S_1$ is $-\frac{5}{9} \log{\frac{5}{9}} - \frac{4}{9} \log{\frac{4}{9}}$, while that for $S_2$ is $-\frac{1}{2} \log{\frac{1}{2}} - \frac{1}{2} \log{\frac{1}{2}}$.

Thus, a lower entropy signifies that a particular sockpuppet is not being used at the same time as the other sockpuppet, while higher entropy indicates that that sockpuppet is being used at the same time as another sockpuppet, with the puppetmaster constantly switching between the two.

Figure~\ref{fig:entropy} shows the entropy of the primary and the secondary sockpuppets in a sockpuppet pair.
We find that secondary accounts tend to have higher entropy, meaning that these sockpuppets are more likely to be used in conjunction with the primary sockpuppet, and thus may be used to support the primary account (e.g., in writing supportive replies).
In contrast, primary sockpuppets have lower entropy, meaning they tend to be used more exclusively.

\begin{figure}
    \centering
    \hspace{-3mm}
    \subfigure[\hspace{-9mm}]{
        \includegraphics[width=0.32\columnwidth]{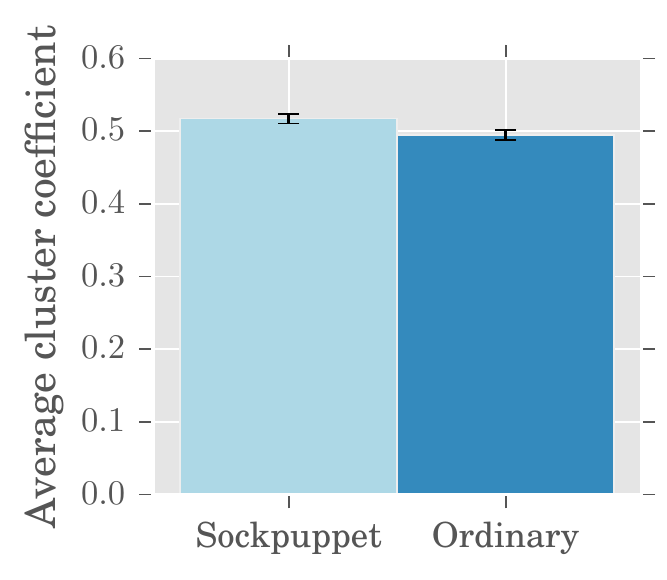}
        \label{fig:reply-cc-local}
    }
    \hspace{-3mm}
    \subfigure[\hspace{-9mm}]{
        \includegraphics[width=0.32\columnwidth]{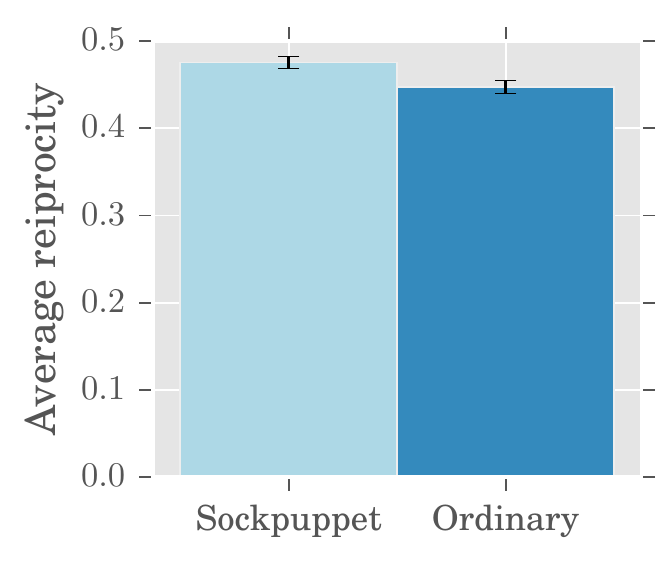}
        \label{fig:reply-avg-reciprocity}
    }
    \hspace{-3mm}
    \subfigure[\hspace{-9mm}]{
        \includegraphics[width=0.35\columnwidth]{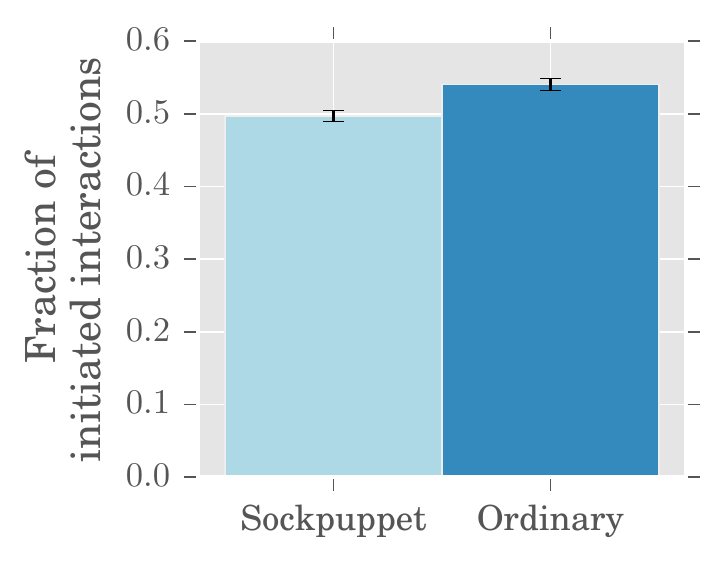}
        \label{fig:reply-incoming}
    }
    \vspace{-3mm}
    \caption{
    Comparison of egonetwork of sockpuppets and similar random users in the reply network.
    }
    \label{fig:network}
    \vspace{-3mm}
\end{figure}

\subsection{Reply network structure}
Last, we examine the user-user interaction network of the entire discussion community. 
To do this, we create a \textit{reply network}, where a node represents a user and an edge from node $A$ to node $B$ indicates that $A$ replied to $B$'s post at least once.
Figure~\ref{fig:avclub} shows the reply network of The AV Club discussion community, with red nodes denoting the sockpuppets and blue nodes denoting ordinary accounts.
Here, we observe that the nodes denoting sockpuppets are more central in the network.
In particular, we find that sockpuppets tend to have higher pagerank than ordinary users (2$\times10^{-4}$ vs 1$\times10^{-6}$, $p$ < 0.001).

To further understand the differences in how the sockpuppets interact with other users, we additionally compare the ego network of sockpuppets with that of ordinary users (Figure~\ref{fig:network}).
We observe that both the number of nodes, and density of the ego networks of sockpuppets and ordinary users are similar (291.5 vs. 291.3 nodes, $p$ = 0.97, and densities of 0.24 vs. 0.22, $p$ < 0.01).
However, the ego networks of sockpuppets are more tightly knit, as measured by the average clustering coefficient (Figure~\ref{fig:reply-cc-local}, 0.52 vs 0.49, $p$ < 0.001).
The nodes in a sockpuppet's ego network reply more to their neighbors, as measured by the average reciprocity (Figure~\ref{fig:reply-avg-reciprocity}, 0.48 vs 0.45, $p$ < 0.001) with sockpuppets generally initiating more interactions (that is, they reply to more users than the users that reply to it, Figure~\ref{fig:reply-incoming}, 0.51 vs 0.46, $p$ < 0.001).
These observations suggest that sockpuppets are highly active in their local network, and also generate more activity among the other users.

\begin{figure*}[t]
    \centering
    \hspace{-3mm}
    \subfigure[\hspace{-9mm}]{
        \includegraphics[width=0.2\textwidth]{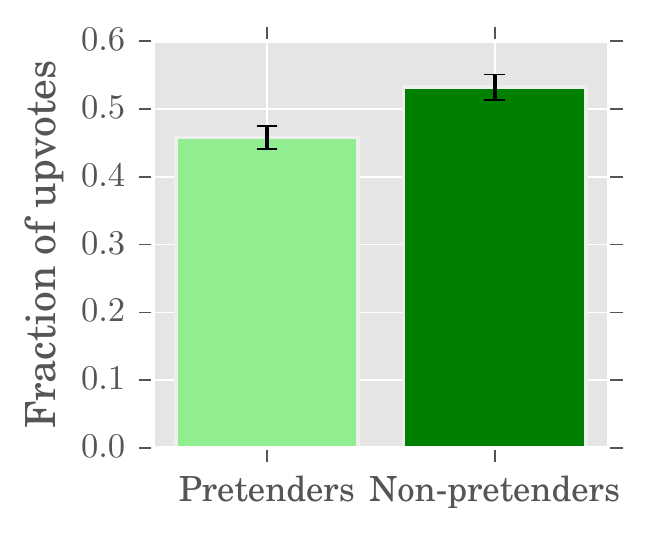}
        \label{fig:x}
    }
    \hspace{-3mm}
    \subfigure[\hspace{-9mm}]{
        \includegraphics[width=0.2\textwidth]{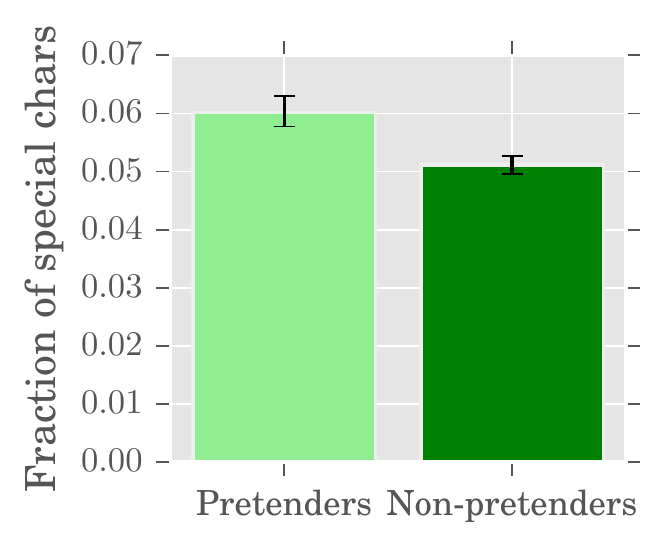}
        \label{fig:x}
    }
    \hspace{-3mm}
    \subfigure[\hspace{-9mm}]{
        \includegraphics[width=0.2\textwidth]{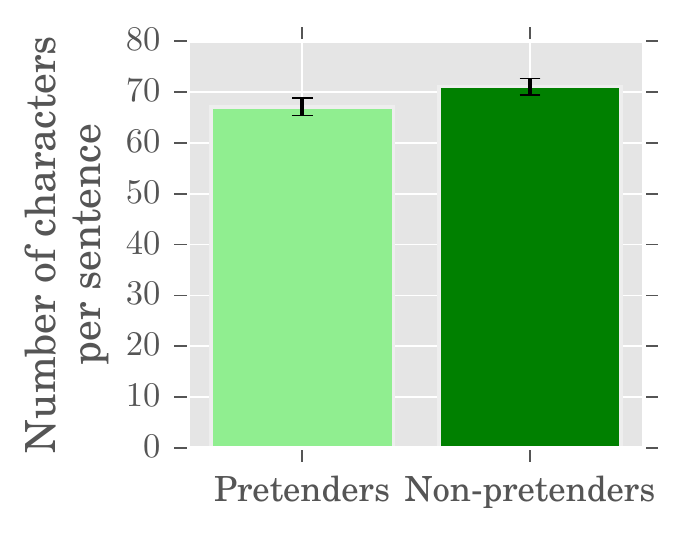}
        \label{fig:x}
    }
    \hspace{-3mm}
    \subfigure[\hspace{-9mm}]{
        \includegraphics[width=0.2\textwidth]{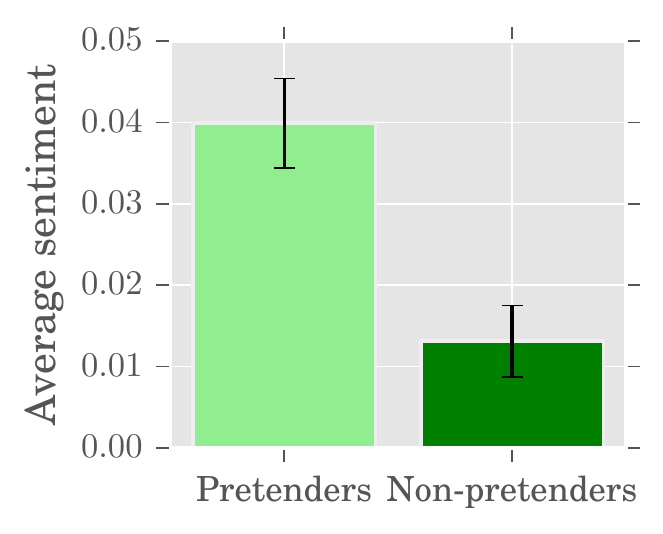}
        \label{fig:x}
    }
    \hspace{-3mm}
    \subfigure[\hspace{-9mm}]{
        \includegraphics[width=0.2\textwidth]{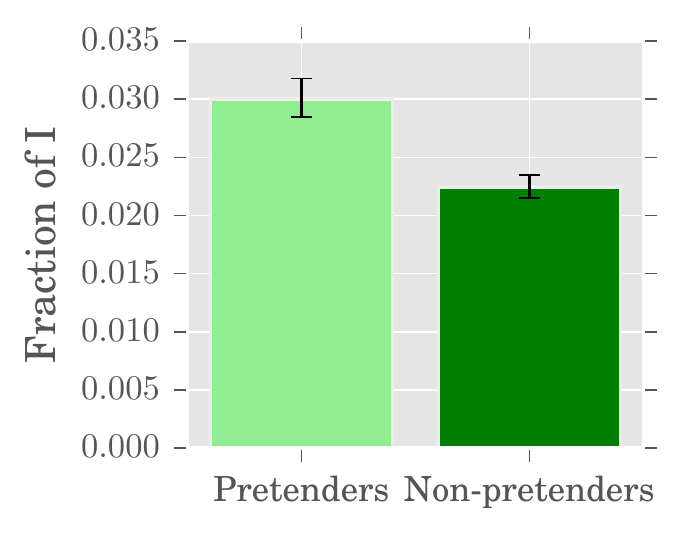}
        \label{fig:x}
    }
    \vspace{-3mm}
    \caption{Differences between pretenders and non-pretenders: (a) fraction of upvotes, (b) fraction of special characters in posts, (c) number of characters per sentence, (d) average sentiment, (e) usage of first person pronoun (``I'').}
    \vspace{-4mm}
    \label{fig:pretenders_stats}
\end{figure*}

\vspace{-2mm}
\section{Types of Sockpuppetry}

Different types of sockpuppets exist, and their characteristics suggest that they may serve different purposes.
Here, in contrast to prior work which assumes that sockpuppets usually pretend to be other users~\cite{solorio2013case, gani2012towards, tsikerdekis2014multiple}, we find that sockpuppets can differ in their deceptiveness -- while many sockpuppets do pretend to be different users, a significant number do not.
When sockpuppets participate in the same discussions, they may also differ in their supportiveness -- sockpuppets may be used to support other sockpuppets of the same puppetmaster, while others may choose not to.

\begin{figure}[t]
    \centering
     \hspace{-3mm}
    \subfigure[\hspace{-9mm}]{
        \includegraphics[width=0.5\columnwidth]{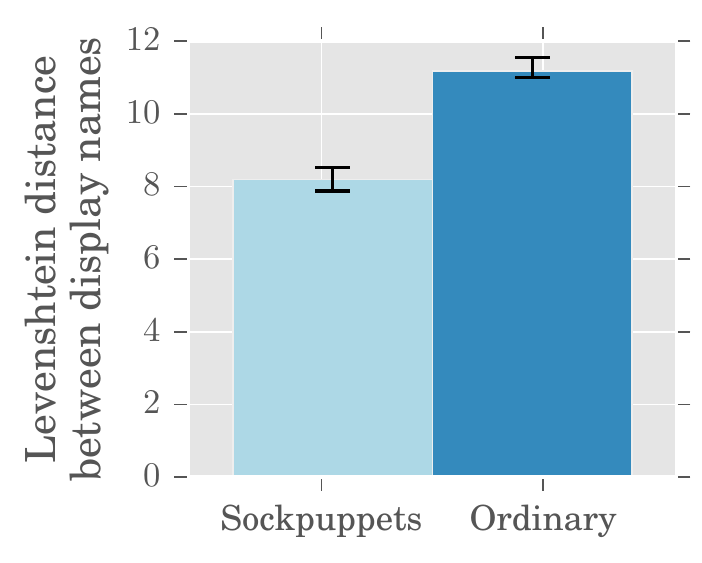}
        \label{fig:usernames}
    }
    \hspace{-3mm}
    \subfigure[\hspace{-9mm}]{
        \includegraphics[width=0.5\columnwidth]{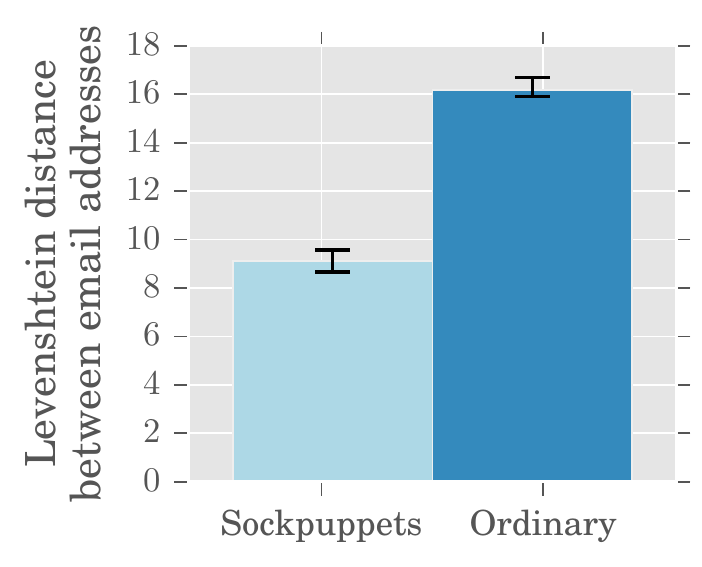}
        \label{fig:useremails}
    }
    \hspace{-3mm}
    \subfigure[\hspace{-9mm}]{
        \includegraphics[width=\columnwidth]{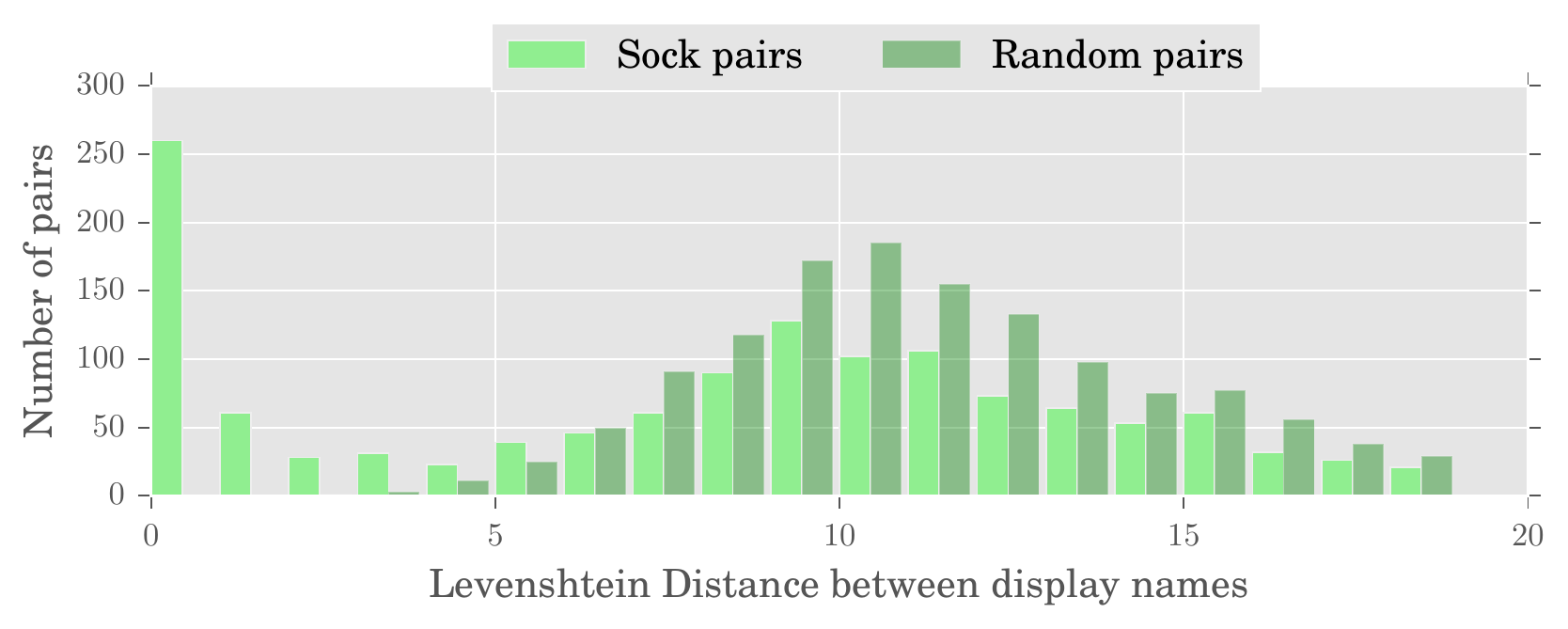}
        \label{fig:pretenders}
    }
    \vspace{-3mm}
    \caption{The (a) display names and (b) email addresses of the sockpuppet accounts are more similar to each other compared to similar random pairs. (c) Based on the distance of display names, sockpuppets can be \textit{pretenders} (high distance) or \textit{non-pretenders} (low distance).}
    \vspace{-6mm}
\end{figure}

\newpage
\vspace{-3mm}
\subsection{Deceptiveness: Pretenders vs. non-pretenders}

A pair of sockpuppets can pretend to be two separate individuals, or may simply be two user accounts an individual uses in different contexts, without any masquerading.
We refer to the former group of sockpuppets as \emph{pretenders}, and the latter group as \emph{non-pretenders}.

One way we might quantify the deceptiveness of a sockpuppet pair is to examine the similarity of display names and email addresses (we only examine the part of the email address before the @-sign).
Display names are public and show up next to user's comments, while email addresses are private and only visible to forum administrators.
If a pair of sockpuppets wants to appear as two separate users, each may adopt a display name that is substantially different from the other in order to deceive community members.
Puppetmaster may also adopt significantly different email addresses to avoid detection by system administrators. To quantify this difference, we measure the Levenshtein distance between two display names, as well as the corresponding email addresses.

\begin{figure}[t]
    \centering
    \hspace{-3mm}
    \subfigure[\hspace{-9mm}]{
        \includegraphics[width=0.48\columnwidth]{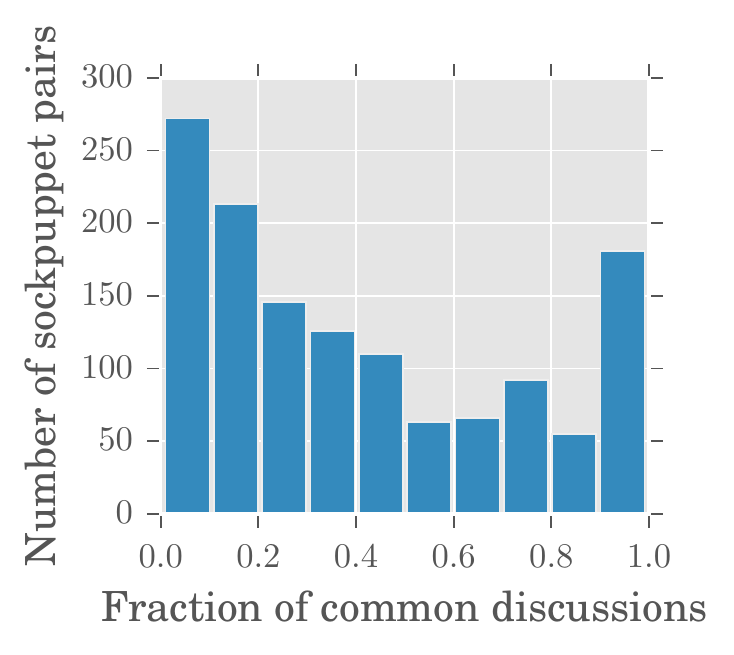}
        \label{fig:frac-common-threads}
    }
    \hspace{-3mm}
    \subfigure[\hspace{-9mm}]{    
        \includegraphics[width=0.52\columnwidth]{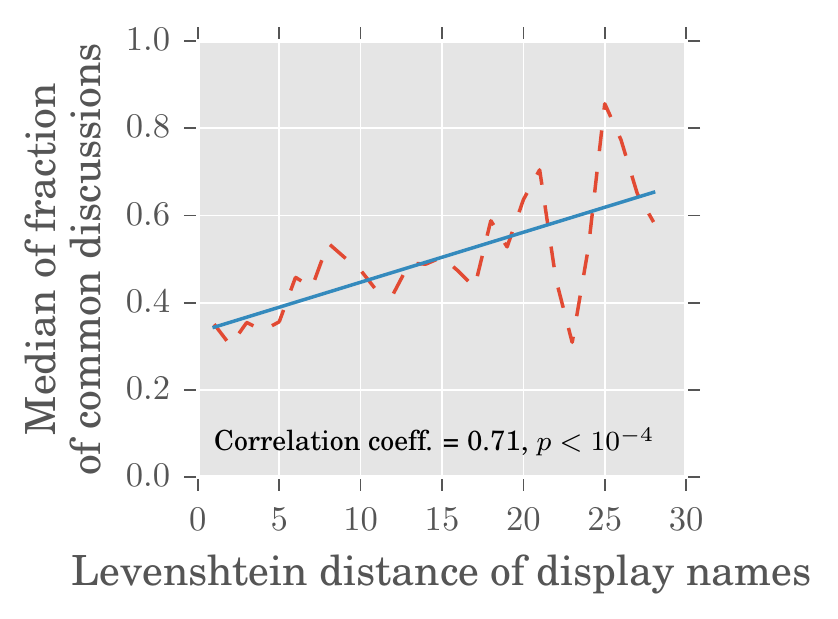}
        \label{fig:pretender-common-threads}
    }
    \vspace{-3mm}
    \caption{(a) Based on the fraction of common discussions between sockpuppet pairs, there are two types of sockpuppets: \textit{independent}, which rarely post on same discussion, and \textit{sock-only}, which only post on same discussions. (b) Increase is display name distance is highly correlated with discussion use.}
    \vspace{-6mm}
 \end{figure}
 
Figures~\ref{fig:usernames} and~\ref{fig:useremails} compare how display names and email addresses differ between pairs of sockpuppets, and between random pairs of ordinary users.
We observe that sockpuppets pairs have both more similar display names as well as email addresses than what would be expected by comparing random pairs of users. This also serves as evidence that sockpuppets we identified are likely to have been created by the same individual.
Further, we also observe that email addresses of sockpuppets are 50\% more similar than those of ordinary accounts, while display names of sockpuppets are only 25\% more similar than expected at random.
This observation may be explained by the fact that sockpuppets put more effort into picking unique display names, which are public-facing, and less effort into picking unique email addresses, which are private-facing and less likely to be noticed.

But are all sockpuppets simply more likely to have more similar display names?
Examining the distribution of the distances between display names in Figure~\ref{fig:pretenders}, we find that the distribution for random pairs is unimodal, while for sockpuppets it is bimodal.
This bimodality suggests that two types of sockpuppets pairs exist. 
The first type of sockpuppets has virtually identical display names (Levenshtein distance $<$ 5), and these are what we call non-pretenders.
The second type of sockpuppets has substantially different display names (Levenshtein distance $\ge$5), and we call them pretenders. Pretenders are likely to be created for deception and use different display names to avoid detection.
Non-pretenders on the other hand have similar display names and this may implicitly signal to the community that they are controlled by the same individual, and thus may be less likely to be malicious.

Across all communities, we find 947 pretender and 403 non-pretender sockpuppet groups. 
We observe that pretenders tend to participate in the same discussions.
For example, Figure~\ref{fig:frac-common-threads} plots the fraction of common discussions over all sockpuppet groups.
We observe bimodality here as well, which may be partially explained by the bimodality of the distribution of display name distances -- 
Figure~\ref{fig:pretender-common-threads} shows that the likelihood of a pair of sockpuppets participating in the same discussion increases as their display names become more different (fitted regression line shown in solid for clarity).
In other words, these observations suggest that sockpuppets that participate in many common discussions have very different display names (high Levenshtein distance) and are thus pretenders, while accounts that participate in few common discussions tend to have similar display names and are thus non-pretenders.

Figure~\ref{fig:pretenders_stats} additionally illustrates the other differences of pretenders and non-pretenders.
We find that pretenders' posts are both more likely to be reported (0.06 fraction of all pretenders' posts are reported vs 0.03 for non-pretenders, $p$ < 0.001 ), be deleted by moderators (0.11 vs 0.08, $p$ < 0.001), and receive a smaller fraction of up-votes (0.45 vs 0.53, $p$ < 0.001).
Pretenders also write posts that contain more uppercase (0.07 vs 0.05, $p$ < 0.001) and special characters (0.06 vs 0.05, $p$ < 0.001), which suggests both shouting, as well as swearing.
In contrast, non-pretenders wrote posts which were longer (35.2 words vs 38.6, $p$ < 0.05) and more readable (ARI 11.15 vs 11.58, $p$ < 0.05).
Pretenders' posts also contained more positive sentiment (0.04 vs 0.013, $p$ < 0.001) and agreement words (0.006 vs 0.005, $p$ < 0.001), suggesting that they tended to be more affable.

\vspace{-2mm}

\subsection{Supporters vs. Dissenters}
Prior work suggests that a primary purpose of sockpuppets is to sway public opinion by creating consensus~\cite{seife2014virtual}.
Thus, we focus our attention on sockpuppets participating in the same discussion, and examine how they interact with each other -- do sockpuppets tend to support each other?

We study two ways in which a sockpuppet pair $(S_1,S_2)$ may interact -- directly, where one sockpuppet ($S_2$) replies to another sockpuppet ($S_1$), or indirectly, where one sockpuppet ($S_2$) replies to a third user ($O$) who had replied to the first sockpuppet ($S_1$). 

We focus on the extent to which sockpuppet $S_2$ agrees with sockpuppet $S_1$, and measure agreement as the difference between fraction of words categorized by LIWC as assenting and those categorized as either negations or dissenting~\cite{yuan2000positive}.
We additionally adjust the sign of agreement depending on who the replying sockpuppet is replying to. For example, $S_2$ may write a post disagreeing with an ordinary user $O$. But if that ordinary user $O$ in turn disagreed with the initial sockpuppet $S_1$, then we assume that the replying sockpuppet $S_2$ is instead in agreement with the initial sockpuppet $S_1$.
We divide sockpuppets into three groups -- \emph{supporters}, who have a positive agreement score, \emph{non-supporters}, who have an agreement score of zero, and \emph{dissenters}, who have a negative agreement score.

Across all communities, we find that 
60\% of the sockpuppets are non-supporters, while 30\% are supporters.
Only 10\% of sockpuppets are dissenters.
Examining these discussions, we find evidence that supporters tend to support the arguments of $S_1$ (e.g., `I agree, or `so true'), and sometimes make additional arguments in their favor (e.g., `That will cost him the election [...]').

On the other hand, dissenters tend to argue against $S_1$ (e.g., `That's not what you said [...]').
We hypothesize that one reason sockpuppets may disagree with each other, despite being controlled by the same puppetmaster, may simply be to attract more attention to the argument.
In some cases, we observed a dissenter making easily refutable arguments (e.g., `Ok if your [sic] so worried about being spied Throw away all your electronics.'), which may have served to discredit the opposing view.

Altogether, these observations suggest that within discussions, sockpuppets may adopt different roles.
While most sockpuppets argue for other sockpuppets controlled by the same puppetmaster, a small but significant number instead argue against other sockpuppets instead.

Nonetheless, is there a relationship between deceptiveness and supportiveness?
Figure \ref{tab:deceptive-supportive} shows that overall, users who support other users in discussions are most likely to be also pretending to be other users (74\% of supporters are pretenders).
Interestingly, when users dissent in a discussion, they are less likely to be a pretender. 
This suggests that pretending is most important when a puppetmaster is trying to create an illusion of consensus.

\begin{table}
\centering

	\begin{tabular}{ccc}
	\hline
	& Pretender & Non-pretender\\ 
	\hline
	Supporter & 0.74 & 0.26  \\
	Non-supporter & 0.70 & 0.30 \\
	Dissenter & 0.58 & 0.42 \\
	\hline
	\end{tabular}
	\caption{74\% of supporters, 70\% of non-supporters and 58\% of dissenters are pretenders. }
    \vspace{-1mm}
	\label{tab:deceptive-supportive}
\end{table}

\vspace{-2mm}

\section{Detecting Sockpuppets}
Our previous analysis found that sockpuppets generally contribute worse content and engage in deceptive behavior.
Thus, it would be useful to create automated tools that can help identify sockpuppets, and assist moderators in policing online communities.
In this section, we consider two classification tasks, both of which relate to the prediction of sockpuppetry.
First, can we distinguish sockpuppets from ordinary users?
And second, can we identify pairs of sockpuppets in the communities?

\begin{table}
\small \begin{tabular}{ll}
\hline
Feature Set & Features \\
\hline
Activity & Reply egonetwork clustering coefficient and reciprocity,\\
& Number of posts, proportion of reply posts, \\
\vspace{1mm}& Time between posts, tenure time \\
%
Community & Whether account is blocked, fraction of upvotes, \\
\vspace{1mm}& Fraction of reported and deleted posts  \\
Post & Number of characters, syllables, words, sentences, \\
& Fraction of punctuations, uppercase characters, etc., \\
& Number of syllables per word, words per sentences, etc. \\
& Readability metrics (\eg\ ARI), LIWC (\eg\ swear words), \\
& Agreement, sentiment and emotion strength \\\hline
\end{tabular}
    \vspace{-2mm}
\caption{Three sets of features were used to identify sockpuppets and sockpuppet pairs.}
    \vspace{-4mm}
\end{table}

Based on the observations and findings from the analyses in the previous sections, we identify three sets of features that may help in finding sockpuppets and sockpuppet pairs: activity features, community features, and post features.
For each user $U$, we develop the following features:

\textbf{Activity features:}
This set of features is derived from $U$'s posting activity.
Prior research has shown that activity behavior of bots, spammers, and vandals is different from that of benign users~\cite{chen2007user, kumar2015vews, kumar2016disinformation, dickerson2014using, tsikerdekis2014multiple}.
Moreover, in our analysis, we have seen that sockpuppets make more posts and they start less sub-discussions. 
Therefore, the activity features we consider include the number of posts, the proportion of posts that are replies, the mean time between two consecutive posts, and $U$'s site tenure, or the number of days from $U$'s first post. 
Further, we use features based on how $U$ is situated in the reply network.
Here, $U$'s local network consists of $U$, the users whose posts $U$ replied to, and the users that replied to $U$'s posts.
We then consider clustering coefficient and reciprocity of this network.
In addition, for the task of identifying pairs of sockpuppets, we use number of common sub-discussions between these sockpuppets to measure how often the two comment together.

\textbf{Community features:} 
Interactions between a user and the rest of the community may also be indicative of sockpuppetry.
Community feedback on an account's posts has been effective in identifying trolls and cheaters~\cite{cheng2015antisocial, blackburn2012branded}, and we also observed that sockpuppets are treated more harshly than ordinary users.
Thus, we consider the fraction of downvotes on posts $U$ wrote, as well as the fraction that were reported or deleted, in addition to whether $U$ was blocked.

\textbf{Post features:}
Finally, we also measure the linguistic features of $U$'s posts.
These features have been very effective to identify sockpuppets~\cite{solorio2013case, bu2013sock}, authors of text~\cite{argamon2005measuring, johansson2013detecting}, deceptive writing styles~\cite{afroz2012detecting}, trolls~\cite{cheng2015antisocial}, hoaxes~\cite{kumar2016disinformation}, and vandalism~\cite{potthast2008automatic}.
In our analysis, we observe that sockpuppets do indeed write differently, for example, writing shorter sentences, using more swear words, and using more singular first-person pronouns.
Thus, we incorporate linguistic features such as the number of characters, average word length, average number of syllables in words, number of big sentences, text readability (measured using ARI), and the different categories of LIWC features.
Finally, we also consider sentiment, emotional valence, and agreement (as described previously).

\begin{figure}
    \centering
    \hspace{-3mm}
    \subfigure[\hspace{-9mm}]{
        \includegraphics[width=0.48\columnwidth]{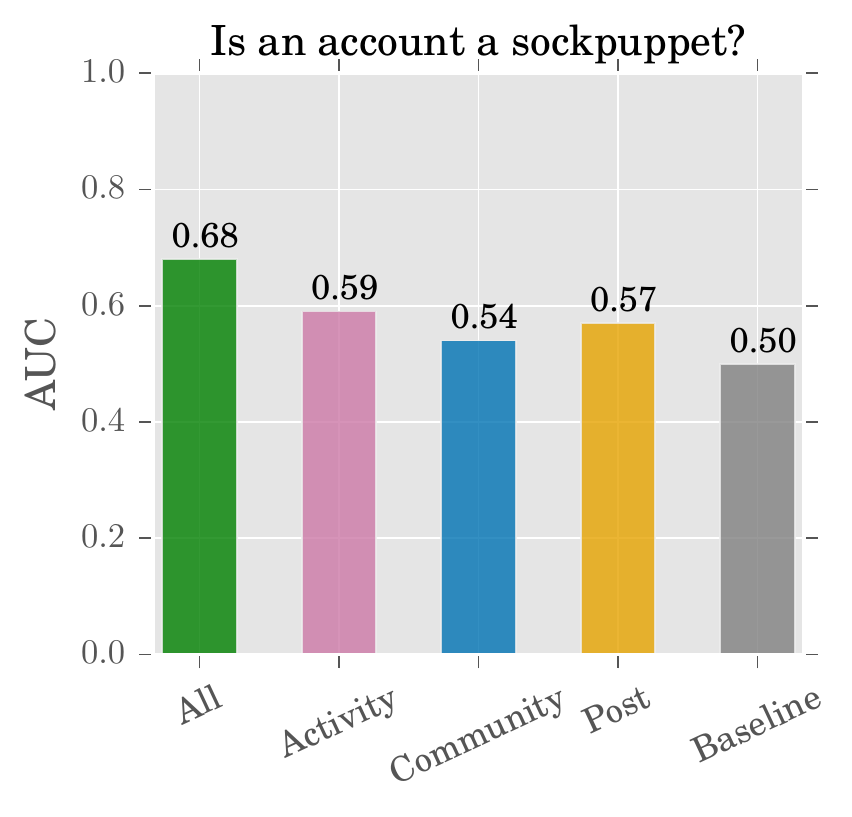}
        \label{fig:task0-auc}
    }
    \hspace{-3mm}
    \subfigure[\hspace{-9mm}]{
        \includegraphics[width=0.48\columnwidth]{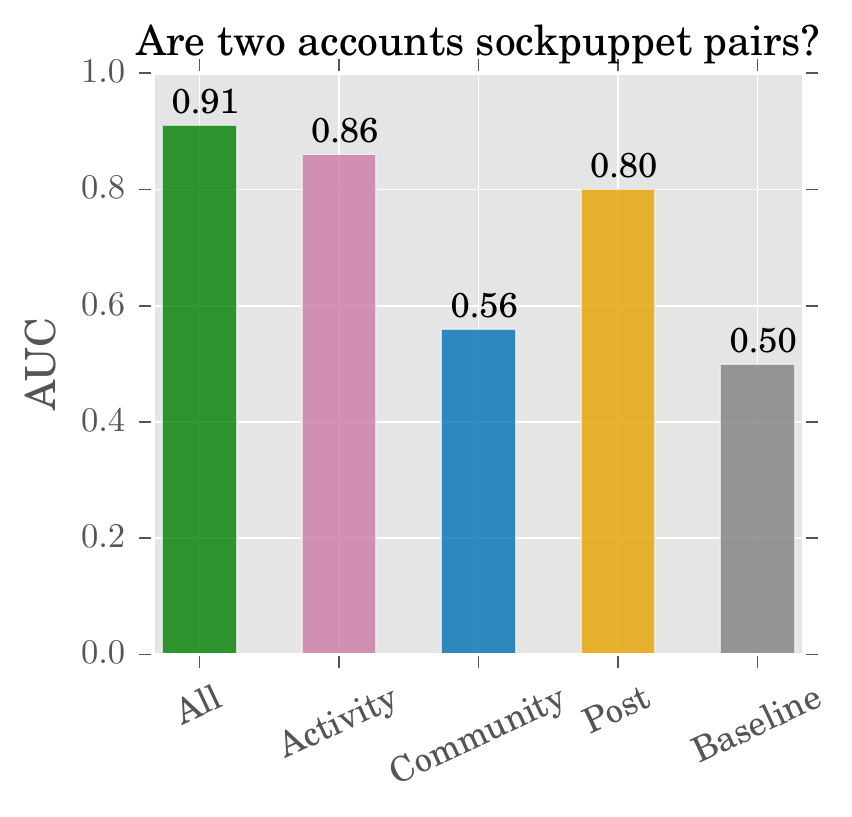}
        \label{fig:task1-auc}
    }
    \vspace{-3mm}
    \caption{Classification performance to identify (a) sockpuppets from ordinary users and (b) pairs of sockpuppet accounts (bottom). Activity features have the highest performance.
    }
    \vspace{-6mm}
 \end{figure}

\vspace{-2mm}
\subsection{Is an account a sockpuppet?}
Given the posts a user has made, is it possible to distinguish a sockpuppet from an ordinary user?
To control for user activity, we match sockpuppets and ordinary users on their total number of posts, as well as the discussions they post in.
Matching gives a balanced dataset, where random guessing results in 50\% accuracy.
We perform 10-fold cross validation using a random forest classifier, then measure performance using ROC AUC.

Figure~\ref{fig:task0-auc} shows results obtained individually with each feature set, as well as when considering all the features together.
We observe that when using all features, the AUC is 0.68.
Individually, all three feature sets perform similarly, with activity features slightly more predictive than the others (AUC=0.59).

To find the relative advantage of adding features, we perform forward feature selection.
We observe that activity and post features perform close to the final AUC of 0.68, and that there is not much lift by adding the community features.
This means that to identify sockpuppets, their activity and content of the post matter the most.

\vspace{-2mm}
\subsection{Are two accounts sockpuppet pairs?}
Next, we turn our attention to identifying pairs of sockpuppets.
Given a sockpuppet and two other users, where one is a sockpuppet, can we predict which user is the sockpuppet?

For each sockpuppet pair $(S_1, S_2)$, we choose a matching ordinary account $O$ for sockpuppet $S_1$. Again, this results in a balanced dataset and the task is to identify which of these two pairs is a sockpuppet. 
Features used in this experiment are the differences in the individual feature values for the two accounts each pair.
We again evaluate performance using a random forest classifier.

Figure~\ref{fig:task1-auc} shows the performance of the resulting classifier.
We achieve a very high AUC of 0.91, suggesting that the interactions between sockpuppets are strongly predictive. 
Looking at the individual features, we see that activity features again perform the best, with close to 0.86 AUC.
Community features perform the poorest, with an AUC of 0.56.

Overall, our results suggest that activity-based features best predict sockpuppetry.
While it is possible to differentiate sockpuppets from ordinary users, it is significantly easier to find other sockpuppets in the same group once at one sockpuppet has been identified.
The latter result suggests most importantly, the interactions between sockpuppets are the best way to identify them.


\vspace{-3mm}

\section{Related Work}
\vspace{-1mm}
Our findings build on a rich vein of prior work in both deception and author identification.

\noindent \textbf{Deception detection}
Sockpuppetry is situated in the broader field of deception. 
Deception online is aided by the virtue of anonymity~\cite{donath1999identity}.
It can occur as deceptive content as well as a deceptive agent~\cite{tsikerdekis2014online}.
The behavior of people changes when they deceive, for example, they reduce communication~\cite{zhou2008cues} and change the focus of their presentation \cite{toma2012lies}.
When writing deceptively to hide their identity, authors tend to increase use of particles and personal pronouns, write shorter sentences, and show nervousness \cite{afroz2012detecting, brennan2012adversarial,juola2012detecting, buller1996interpersonal}.
Our work adds to this line of research by finding evidence of deceptive writing styles and presentation by sockpuppets -- pretender sockpuppets may pretend to be different people by using different display names and they tend to write deceptively as well. 

\noindent \textbf{Motivations for sockpuppetry}
Turning to research that studies sockpuppetry specifically, one line of work has studied their motivations.
Sockpuppetry is often used to avoid being banned, to create false consensus \cite{stone2007hand,afroz2012detecting} and support a person or a position \cite{caspi2006online}, or vandalize content (\eg\ on Wikipedia \cite{solorio2013case}).
Relatedly, motivations for multiple account creation in online multiplayer games can either be benign (\eg\ experimentation with different identities) or malicious (\eg\ increasing in-game profit, cheating)~\cite{gilbert2014psychological, gilbert2011multiple, kafai2007your, caspi2006online}.
In our work, we find evidence for these motivations -- sockpuppets in discussion communities sometimes support each other, and beyond malicious uses, some uses of sockpuppetry may be benign (\eg\ a user may simply use different accounts to post in different topics).

\noindent \textbf{Sockpuppetry and author identification}
Another line of work has also identified sockpuppets using textual information, link analysis and temporal information, both in online discussion forums~\cite{bu2013sock, zheng2011sockpuppet} and social networks~\cite{gani2012towards, liu2016sockpuppet, zheng2011sockpuppet,viswanath2010analysis}.
However, definitions of sockpuppets from previous research have tended to make assumptions about the usernames that sockpuppets use (\eg, that they are similar \cite{liu2016sockpuppet}), their opinion towards topics (\eg, they have the same opinion \cite{bu2013sock}), 
and their interactions (e.g., that they reply in support of each other's posts \cite{zheng2011sockpuppet}).
As such, these definitions tend to miss several types of sockpuppetry.
In this work, we developed a robust methodology for identifying sockpuppets that makes fewer assumptions, and showed that a significant fraction of sockpuppets do use different names (i.e., the non-pretenders), and tend not to support each other in discussions (i.e., the non-supporters).

Sockpuppetry on Wikipedia has been studied extensively due to availability of manually-validated ground-truth data~\cite{solorio2013case, tsikerdekis2014multiple, yamak2016detection, paul2015editing}.
However, in contrast to sockpuppetry in discussion communities, which is the focus of our work, sockpuppet editors on Wikipedia primarily edit articles, and the main purpose is not to interact with each other.

Closely related to sockpuppet detection is author identification, or the task of identifying the original author of a document~\cite{johansson2013detecting, johansson2015timeprints, narayanan2012feasibility, novak2004anti, paul2015editing, qian2013identifying}.
More recently, research has used multiple accounts of users across different social platforms to identify malicious users~\cite{venkatadri2016strengthening}, and to identify accounts operated by the same user across different web platforms~\cite{jain2013seek, silvestri2015linking, zafarani2013connecting, zheng2011sockpuppet}.
In contrast to our work here, this line of research does not operate under the assumption of deception and thus may be less applicable to situations when authors try to obfuscate their writing~\cite{afroz2012detecting, brennan2012adversarial}. 


\section{Discussion and Conclusion}

Our findings shed light on how sockpuppets are used in practice in online discussion communities.
By developing a robust methodology for identifying sockpuppets, we are able to comprehensively study their activity.
Importantly, this methodology is able to identify sockpuppets that were created at significantly different times, use very different usernames or email addresses, write differently, or mostly post in different discussions.

Our work revealed differences in how sockpuppets write and behave in online communities.
Sockpuppets use more singular first-person pronouns, write shorter sentences, and swear more.
They participate in discussions with more controversial topics, and are especially likely to interact with other sockpuppets.
These differences allowed us to build preditive models that robustly differentiate pairs of sockpuppets from ordinary users, as well as identify individual user accounts that are sockpuppets.

Nonetheless, our analysis has limitations that would be interesting to explore in future work.
First, using our heuristics, we are not able to identify sockpuppets that are also throwaway accounts and are only used once before being abandoned. 
Next, we studied sockpuppetry in discussion forums where users are pseudonymous.
It would be interesting to study 
the effect of using real identities (e.g., Facebook), or in completely anonymous settings (e.g., 4chan).
We also primarily studied pairs of sockpuppets, but understanding how larger groups of sockpuppets function may reveal additional ways in which sockpuppets may coordinate, and may allow us to observe more pronounced effects of sockpuppetry.
Further, behavior of sockpuppets in knowledge sharing platforms (e.g., StackOverflow) may be different from that in opinion expressing discussion platforms we studied -- for instance, the primary purpose of sockpuppets in such platforms may primarily be to give additional `upvotes' to their answers. Such a study would bring additional insights into sockpuppetry.
Furthermore, prior work found that trolling correlates with sadism \cite{buckels2014trolls} -- understanding the role of personality traits in sockpuppetry would also be valuable future work.
Finally, even more robust methodologies for identifying sockpuppets may uncover an even wider range of behavior.
For example, sockpuppets may exist beyond a single discussion community -- for example, we found 14 different sockpuppet groups that were used in more than one online community. Studying these types of sockpuppets may allow us to better characterize how a sockpuppet's behavior may change in different communities.

By developing better techniques to identify sockpuppets, our work can be used to improve the quality of discussions online, where each discussant can better trust that their interactions with others are genuine.
Nonetheless, while it is possible to identify sockpuppetry, one should be careful not to assume that all sockpuppets are malicious.
Our work suggests that a significant number of sockpuppets do not pretend to be other users and were simply used to participate in different discussions.
This observation suggests that some users find it valuable to separate their activity in different spheres of interests.

\section{Acknowledgement}
Parts of this work were supported by US Army Research Office under Grant Number W911NF1610342, NSF IIS-1149837, ARO MURI, DARPA NGS2, Stanford Data Science Initiative and Microsoft Research PhD fellowship. We would like to thank Disqus for sharing data with us for research and the anonymous reviewers for their helpful comments.

\newpage
\balance
\bibliographystyle{abbrv}
\bibliography{bibliography}

\begin{thebibliography}{10}

\bibitem{wikisocks}
Wikipedia sockpuppet investigation policy.
\newblock \url{https://goo.gl/89ieoY}.
\newblock Accessed: 2016-10-24.

\bibitem{afroz2012detecting}
S.~Afroz, M.~Brennan, and R.~Greenstadt.
\newblock Detecting hoaxes, frauds, and deception in writing style online.
\newblock In {\em Security and Privacy (SP), 2012 IEEE Symposium on}, pages
  461--475. IEEE, 2012.

\bibitem{argamon2005measuring}
S.~Argamon and S.~Levitan.
\newblock Measuring the usefulness of function words for authorship
  attribution.
\newblock In {\em Proceedings of the 2005 ACH/ALLC Conference}, 2005.

\bibitem{blackburn2012branded}
J.~Blackburn, R.~Simha, N.~Kourtellis, X.~Zuo, M.~Ripeanu, J.~Skvoretz, and
  A.~Iamnitchi.
\newblock Branded with a scarlet c: cheaters in a gaming social network.
\newblock In {\em Proceedings of the 21st international conference on World
  Wide Web}, pages 81--90. ACM, 2012.

\bibitem{brennan2012adversarial}
M.~Brennan, S.~Afroz, and R.~Greenstadt.
\newblock Adversarial stylometry: Circumventing authorship recognition to
  preserve privacy and anonymity.
\newblock {\em ACM Transactions on Information and System Security (TISSEC)},
  15(3):12, 2012.

\bibitem{bu2013sock}
Z.~Bu, Z.~Xia, and J.~Wang.
\newblock A sock puppet detection algorithm on virtual spaces.
\newblock {\em Knowledge-Based Systems}, 37:366--377, 2013.

\bibitem{buckels2014trolls}
E.~E. Buckels, P.~D. Trapnell, and D.~L. Paulhus.
\newblock Trolls just want to have fun.
\newblock {\em Personality and individual Differences}, 67:97--102, 2014.

\bibitem{buller1996interpersonal}
D.~B. Buller and J.~K. Burgoon.
\newblock Interpersonal deception theory.
\newblock {\em Communication theory}, 6(3):203--242, 1996.

\bibitem{caspi2006online}
A.~Caspi and P.~Gorsky.
\newblock Online deception: Prevalence, motivation, and emotion.
\newblock {\em CyberPsychology \& Behavior}, 9(1):54--59, 2006.

\bibitem{chen2007user}
K.-T. Chen and L.-W. Hong.
\newblock User identification based on game-play activity patterns.
\newblock In {\em Proceedings of the 6th ACM SIGCOMM workshop on Network and
  system support for games}, pages 7--12. ACM, 2007.

\bibitem{cheng2015antisocial}
J.~Cheng, C.~Danescu-Niculescu-Mizil, and J.~Leskovec.
\newblock Antisocial behavior in online discussion communities.
\newblock In {\em Ninth International AAAI Conference on Web and Social Media},
  2015.

\bibitem{dickerson2014using}
J.~P. Dickerson, V.~Kagan, and V.~Subrahmanian.
\newblock Using sentiment to detect bots on twitter: Are humans more
  opinionated than bots?
\newblock In {\em Advances in Social Networks Analysis and Mining (ASONAM),
  2014 IEEE/ACM International Conference on}, pages 620--627. IEEE, 2014.

\bibitem{donath1999identity}
J.~S. Donath et~al.
\newblock Identity and deception in the virtual community.
\newblock {\em Communities in cyberspace}, 1999.

\bibitem{gani2012towards}
K.~Gani, H.~Hacid, and R.~Skraba.
\newblock Towards multiple identity detection in social networks.
\newblock In {\em Proceedings of the 21st International Conference on World
  Wide Web}, pages 503--504. ACM, 2012.

\bibitem{gilbert2014psychological}
R.~Gilbert, V.~Thadani, C.~Handy, H.~Andrews, T.~Sguigna, A.~Sasso, and
  S.~Payne.
\newblock The psychological functions of avatars and alt (s): A qualitative
  study.
\newblock {\em Computers in Human Behavior}, 32:1--8, 2014.

\bibitem{gilbert2011multiple}
R.~L. Gilbert, J.~A. Foss, and N.~A. Murphy.
\newblock Multiple personality order: Physical and personality characteristics
  of the self, primary avatar and alt.
\newblock In {\em Reinventing ourselves: Contemporary concepts of identity in
  virtual worlds}, pages 213--234. Springer, 2011.

\bibitem{hancock2007digital}
J.~T. Hancock.
\newblock Digital deception.
\newblock {\em Oxford handbook of internet psychology}, pages 289--301, 2007.

\bibitem{vader}
C.~J. Hutto and E.~Gilbert.
\newblock Vader: A parsimonious rule-based model for sentiment analysis of
  social media text.
\newblock In {\em Eighth International AAAI Conference on Weblogs and Social
  Media}, 2014.

\bibitem{jain2013seek}
P.~Jain, P.~Kumaraguru, and A.~Joshi.
\newblock @ i seek'fb. me': Identifying users across multiple online social
  networks.
\newblock In {\em Proceedings of the 22nd international conference on World
  Wide Web}, pages 1259--1268. ACM, 2013.

\bibitem{johansson2013detecting}
F.~Johansson, L.~Kaati, and A.~Shrestha.
\newblock Detecting multiple aliases in social media.
\newblock In {\em Proceedings of the 2013 IEEE/ACM international conference on
  advances in social networks analysis and mining}, pages 1004--1011. ACM,
  2013.

\bibitem{johansson2015timeprints}
F.~Johansson, L.~Kaati, and A.~Shrestha.
\newblock Timeprints for identifying social media users with multiple aliases.
\newblock {\em Security Informatics}, 4(1):7, 2015.

\bibitem{juola2012detecting}
P.~Juola.
\newblock Detecting stylistic deception.
\newblock In {\em Proceedings of the Workshop on Computational Approaches to
  Deception Detection}, pages 91--96. Association for Computational
  Linguistics, 2012.

\bibitem{kafai2007your}
Y.~B. Kafai, D.~A. Fields, and M.~Cook.
\newblock Your second selves: avatar designs and identity play in a teen
  virtual world.
\newblock In {\em Proceedings of DIGRA}, volume 2007, 2007.

\bibitem{kumar2015vews}
S.~Kumar, F.~Spezzano, and V.~Subrahmanian.
\newblock Vews: A wikipedia vandal early warning system.
\newblock In {\em Proceedings of the 21th ACM SIGKDD International Conference
  on Knowledge Discovery and Data Mining}, pages 607--616. ACM, 2015.

\bibitem{kumar2016disinformation}
S.~Kumar, R.~West, and J.~Leskovec.
\newblock Disinformation on the web: Impact, characteristics, and detection of
  wikipedia hoaxes.
\newblock In {\em Proceedings of the 25th International Conference on World
  Wide Web}, pages 591--602. International World Wide Web Conferences Steering
  Committee, 2016.

\bibitem{liu2016sockpuppet}
D.~Liu, Q.~Wu, W.~Han, and B.~Zhou.
\newblock Sockpuppet gang detection on social media sites.
\newblock {\em Frontiers of Computer Science}, 10(1):124--135, 2016.

\bibitem{mukherjee2013yelp}
A.~Mukherjee, V.~Venkataraman, B.~Liu, and N.~Glance.
\newblock What yelp fake review filter might be doing?
\newblock In {\em Seventh International AAAI Conference on Weblogs and Social
  Media}, 2013.

\bibitem{narayanan2012feasibility}
A.~Narayanan, H.~Paskov, N.~Z. Gong, J.~Bethencourt, E.~Stefanov, E.~C.~R.
  Shin, and D.~Song.
\newblock On the feasibility of internet-scale author identification.
\newblock In {\em Security and Privacy (SP), 2012 IEEE Symposium on}, pages
  300--314. IEEE, 2012.

\bibitem{novak2004anti}
J.~Novak, P.~Raghavan, and A.~Tomkins.
\newblock Anti-aliasing on the web.
\newblock In {\em Proceedings of the 13th international conference on World
  Wide Web}, pages 30--39. ACM, 2004.

\bibitem{paul2015editing}
P.~P. Paul, M.~Sultana, S.~A. Matei, and M.~Gavrilova.
\newblock Editing behavior to recognize authors of crowdsourced content.
\newblock In {\em Systems, Man, and Cybernetics (SMC), 2015 IEEE International
  Conference on}, pages 1676--1681. IEEE, 2015.

\bibitem{pennebaker2001linguistic}
J.~W. Pennebaker, M.~E. Francis, and R.~J. Booth.
\newblock Linguistic inquiry and word count: Liwc 2001.
\newblock {\em Mahway: Lawrence Erlbaum Associates}, 71(2001):2001, 2001.

\bibitem{potthast2008automatic}
M.~Potthast, B.~Stein, and R.~Gerling.
\newblock Automatic vandalism detection in wikipedia.
\newblock In {\em European Conference on Information Retrieval}, pages
  663--668. Springer, 2008.

\bibitem{qian2013identifying}
T.~Qian and B.~Liu.
\newblock Identifying multiple userids of the same author.
\newblock In {\em EMNLP}, pages 1124--1135, 2013.

\bibitem{rosenbaum1983central}
P.~R. Rosenbaum and D.~B. Rubin.
\newblock The central role of the propensity score in observational studies for
  causal effects.
\newblock {\em Biometrika}, pages 41--55, 1983.

\bibitem{seife2014virtual}
C.~Seife.
\newblock {\em Virtual Unreality: Just Because the Internet Told You, how Do
  You Know It's True?}
\newblock Penguin, 2014.

\bibitem{silvestri2015linking}
G.~Silvestri, J.~Yang, A.~Bozzon, and A.~Tagarelli.
\newblock Linking accounts across social networks: the case of stackoverflow,
  github and twitter.
\newblock In {\em International Workshop on Knowledge Discovery on the WEB},
  pages 41--52, 2015.

\bibitem{solorio2013case}
T.~Solorio, R.~Hasan, and M.~Mizan.
\newblock A case study of sockpuppet detection in wikipedia.
\newblock In {\em Workshop on Language Analysis in Social Media (LASM) at NAACL
  HLT}, pages 59--68, 2013.

\bibitem{stone2007hand}
B.~Stone and M.~Richtel.
\newblock The hand that controls the sock puppet could get slapped.
\newblock {\em New York Times}, 2007.

\bibitem{toma2012lies}
C.~L. Toma and J.~T. Hancock.
\newblock What lies beneath: The linguistic traces of deception in online
  dating profiles.
\newblock {\em Journal of Communication}, 62(1):78--97, 2012.

\bibitem{tsikerdekis2014multiple}
M.~Tsikerdekis and S.~Zeadally.
\newblock Multiple account identity deception detection in social media using
  nonverbal behavior.
\newblock {\em IEEE Transactions on Information Forensics and Security},
  9(8):1311--1321, 2014.

\bibitem{tsikerdekis2014online}
M.~Tsikerdekis and S.~Zeadally.
\newblock Online deception in social media.
\newblock {\em Communications of the ACM}, 57(9):72--80, 2014.

\bibitem{venkatadri2016strengthening}
G.~Venkatadri, O.~Goga, C.~Zhong, B.~Viswanath, K.~P. Gummadi, and N.~Sastry.
\newblock Strengthening weak identities through inter-domain trust transfer.
\newblock In {\em Proceedings of the 25th International Conference on World
  Wide Web}, pages 1249--1259. International World Wide Web Conferences
  Steering Committee, 2016.

\bibitem{viswanath2010analysis}
B.~Viswanath, A.~Post, K.~P. Gummadi, and A.~Mislove.
\newblock An analysis of social network-based sybil defenses.
\newblock {\em ACM SIGCOMM Computer Communication Review}, 40(4):363--374,
  2010.

\bibitem{yamak2016detection}
Z.~Yamak, J.~Saunier, and L.~Vercouter.
\newblock Detection of multiple identity manipulation in collaborative
  projects.
\newblock In {\em Proceedings of the 25th International Conference Companion on
  World Wide Web}, pages 955--960. International World Wide Web Conferences
  Steering Committee, 2016.

\bibitem{yuan2000positive}
H.~Yuan, D.~O. Clifton, P.~Stone, and H.~H. Blumberg.
\newblock Positive and negative words: Their association with leadership talent
  and effectiveness.
\newblock {\em The Psychologist-Manager Journal}, 4(2):199, 2000.

\bibitem{zafarani2013connecting}
R.~Zafarani and H.~Liu.
\newblock Connecting users across social media sites: a behavioral-modeling
  approach.
\newblock In {\em Proceedings of the 19th ACM SIGKDD international conference
  on Knowledge discovery and data mining}, pages 41--49. ACM, 2013.

\bibitem{zheng2011sockpuppet}
X.~Zheng, Y.~M. Lai, K.-P. Chow, L.~C. Hui, and S.-M. Yiu.
\newblock Sockpuppet detection in online discussion forums.
\newblock In {\em Intelligent Information Hiding and Multimedia Signal
  Processing (IIH-MSP), 2011 Seventh International Conference on}, pages
  374--377. IEEE, 2011.

\bibitem{zhou2008cues}
L.~Zhou and Y.-w. Sung.
\newblock Cues to deception in online chinese groups.
\newblock In {\em Hawaii international conference on system sciences,
  proceedings of the 41st annual}, pages 146--146. IEEE, 2008.

\end{thebibliography}

\end{document}